\documentclass[twocolumn]{aastex631}
\usepackage{amsmath}
\usepackage[flushleft]{threeparttable}

\shorttitle{Polar flux Redistribution}
\shortauthors{Schonfeld et al.}

\begin{document}

\title{Solar Polar Flux Redistribution based on Observed Coronal Holes}

\correspondingauthor{Samuel J. Schonfeld}
\email{schonfsj@gmail.com}

\author[0000-0002-5476-2794]{Samuel J. Schonfeld}
\affiliation{Institute for Scientific Research, Boston College \\
Kenny Cottle Hall 207A, 885 Centre St. \\
Newton, MA 02459, USA}
\affiliation{Air Force Research Laboratory, Space Vehicles Directorate \\
Kirtland AFB, NM 87117, USA}

\author[0000-0002-6038-6369]{Carl J. Henney}
\affiliation{Air Force Research Laboratory, Space Vehicles Directorate \\
Kirtland AFB, NM 87117, USA}

\author[0000-0001-9498-460X]{Shaela I. Jones}
\affiliation{NASA Goddard Space Flight Center \\
Greenbelt, MD 20771, USA}
\affiliation{Department of Physics, Catholic University of America \\
Washington, DC 20064, USA}

\author[0000-0001-9326-3448]{Charles N. Arge}
\affiliation{NASA Goddard Space Flight Center \\
Greenbelt, MD 20771, USA}



\begin{abstract}
We explore the use of observed polar coronal holes (CHs) to constrain the flux distribution within the polar regions of global solar magnetic field maps in the absence of reliable quality polar field observations.  Global magnetic maps, generated by the Air Force Data Assimilative Photospheric flux Transport (ADAPT) model, are modified to enforce field unipolarity thresholds both within and outside observed CH boundaries. The polar modified and unmodified maps are used to drive Wang-Sheeley-Arge (WSA) models of the corona and solar wind (SW). The WSA predicted CHs are compared with the observations, and SW predictions at the WIND and Ulysses spacecraft are also used to provide context for the new polar modified maps. We find that modifications of the polar flux never worsen and typically improve both the CH and SW predictions. We also confirm the importance of the choice of the domain over which WSA generates the coronal magnetic field solution but find that solutions optimized for one location in the heliosphere can worsen predictions at other locations. Finally, we investigate the importance of low-latitude (i.e., active region) magnetic fields in setting the boundary of polar CHs, determining that they have at least as much impact as the polar fields themselves.
\end{abstract}



\section{Introduction} \label{sec:intro}
\label{sec:introduction}
The solar photospheric magnetic field is the driving boundary condition for solar irradiance variability \citep[e.g.,][]{Kopp2021, Petrie2021}, the solar atmosphere \citep[e.g.,][]{Mackay2012, Wiegelmann2014}, the solar wind \citep[SW; e.g.,][]{Cranmer2019, Rouillard2021}, and the heliosphere \citep[e.g.,][]{Owens2013, Wiegelmann2021}. Models of the inner heliosphere \citep[e.g.,][]{Riley2001, Linker2013, vanderHolst2014} rely on a global ($4\pi$ steradian) photospheric magnetic field inner boundary to extrapolate the field out into the heliosphere. Currently, magnetograms are only available along the Sun-Earth line, and that single line-of-sight results in quality measurements for approximately 1/4\textsuperscript{th} of the solar surface at a time. While solar rotation allows $360\degr$ of longitude to be observed from Earth over approximately 27~days, it does not enable continuous observations of the poles. The $7.23\degr$ offset between the ecliptic and the solar rotation axis (the solar $b$ angle) can be exploited to provide high-resolution polar vector magnetic field measurements \citep{Tsuneta2008b, Petrie2017}, but these are only available at each pole for a short period once per year \citep{Petrie2015}.

Since the magnetic field distribution at the poles is usually a key driver of the dipole component of the solar magnetic field, particularly at solar minimum, the polar observational gap creates significant difficulties for global coronal models. A number of methods have been developed to fill the polar magnetic field directly from lower-latitude observations \citep[e.g.,][]{Sun2011, Linker2013, Linker2017, Sun2018preprint, Mikic2018} and it is also possible to estimate the polar magnetic fields with flux transport models \citep{Worden2000, Schrijver2003, Upton2014}. In particular, the Air Force Data Assimilative Photospheric flux Transport (ADAPT) model \citep{Arge2010a, Arge2011a, Hickmann2015} creates an ensemble of realizations of the photospheric field, each with a unique representation of the polar fields, estimating the uncertainty due to the lack of direct observations \citep[e.g., see Figure 11 in][]{Posner2021}.

Besides full-disk magnetograms, extreme ultraviolet (EUV) observations of polar coronal holes \citep[CHs; e.g.,][]{Burton1968, Withbroe1971, Munro1972, Zirker1977} also provide constraints on the polar magnetic field. CHs are created by concentrations of predominantly ``open'' magnetic fields that appear dark in EUV observations \citep[e.g.,][]{Cranmer2009, Hofmeister2019}, and polar CHs are highly unipolar \citep{Harvey2002, Henney2005}. This strong connection between the coronal magnetic field and CHs makes them an obvious choice to exploit when direct photospheric magnetic field observations are not available. To that end, recent work by \citet{Heinemann2021b} used CH observations to estimate open magnetic flux on the farside of the Sun without magnetic field observations.

Importantly, CHs are also one of the primary sources of the SW \citep[e.g.,][]{Krieger1973, Cranmer2009}. High-speed wind originates from the centers of large CHs while slower, high Alfv\'enicity wind likely originates from closer to the edges of CHs near to the closed-field boundary \citep[e.g.,][]{DAmicis2019, Stansby2019, Wang2019}. This relationship can be described by relating the expansion of the open magnetic field with altitude to the resulting SW speed \citep[e.g.,][]{Withbroe1988, Wang1990}, although the precise physical mechanism responsible for this explanation is not yet known. There is also significant slow wind near the heliospheric current sheet \citep{Smith2001} related to various forms of magnetic reconnection at the open-closed field boundary and from the cusps of helmet streamers \citep[e.g.,][and references therein]{Viall2020a, Rouillard2021}. Consequently, models of the SW are typically quite sensitive to the details of the boundary between open and closed magnetic fields \citep{Riley2015, Wallace2020}.

In this paper, we investigate the potential of using CH observations to constrain polar magnetic fields within global magnetic maps used as input to corona and SW models. We use global photospheric magnetic field maps generated with ADAPT and modify the polar regions in these maps to conform to polar CH observations. The redistributed flux maps are then used to drive Wang-Sheeley-Arge \citep[WSA;][]{Arge2000, Arge2003, Arge2004a, McGregor2008, Wallace2020} models, and we compare the modeled CH maps as well as predictions of the SW speed near the Earth and at high heliospheric latitudes with corresponding observations. The data used in this study are described in section~\ref{sec:data}. We outline the ADAPT model as well as the modifications made to the polar magnetic fields in section~\ref{sec:ADAPT}. The WSA models and comparisons with the observations are related in section \ref{sec:WSA}. We conclude and outline the importance of this work in light of near-future observational capabilities in section~\ref{sec:conclusion}.

\section{Data}
\label{sec:data}
For this preliminary study of applying solar polar magnetic flux redistribution based on polar CH observations, we investigate the period between 1995 September 29\textsuperscript{th} and October 27\textsuperscript{th} (i.e., Carrington rotation 1901, or simply CR1901). This rotation occurred during the activity minimum following solar cycle 22 \citep{Hathaway2015} with an (Earth orbital distance) adjusted F\textsubscript{10.7} in the range 69.8--91.8~sfu \citep{Covington1947, Covington1969, Tapping1987, Tapping2013}. The inclination of the solar rotation axis to the ecliptic (the $b$ angle as observed from Earth) during this period was $6.8\degr$--$4.9\degr$.

\subsection{ADAPT Magnetic Field Maps}
\label{sec:data:ADAPT}
\begin{figure*}[!th]
    \includegraphics[width=\textwidth]{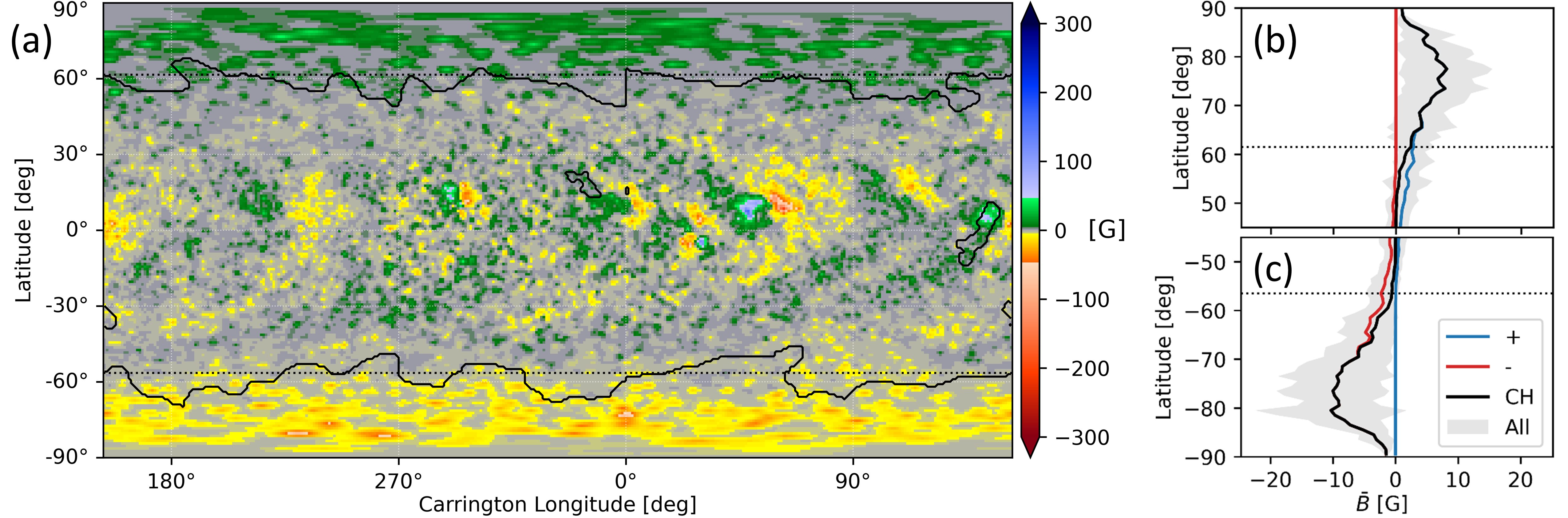}
    \caption{The first realization of the ADAPT ensemble on 1995 October 1\textsuperscript{st}. The map shows the radial photospheric magnetic field and the solid black contours mark the observed coronal hole (CH) boundaries from \protect\citet{Harvey2002}. The line plots indicate the average magnetic field strength as a function of latitude in the north and south poles for the positive (blue), negative (red), and CH (black) flux, along with the standard deviation of all fluxes (shaded gray). In all panels, the horizontal dotted black lines indicate the ADAPT data assimilation range at $-56\degr$ and $+61\degr$.
    \label{fig:ADAPT_original}}
\end{figure*}
The Air Force Data Assimilative Photospheric flux Transport (ADAPT) model \citep{Arge2010a, Arge2011a} creates synchronic global maps \citep[e.g.,][]{Riley2014} of the photospheric radial magnetic field by combining data assimilation \citep{Hickmann2015} with the flux transport scheme of \citet{Worden2000}. Following \citet{Worden2000}, the flux transport of the ADAPT model includes differential rotation, meridional circulation, and supergranular diffusion, along with random weak field flux emergence. To account for uncertainties in the transport processes (e.g., supergranulation results in random flow patterns), ADAPT includes an ensemble of 16 realizations. Four of the model realizations have enhanced flux values that yield non-physical maps but are needed for the current data assimilation method \citep{Hickmann2015}. These realizations are excluded from the public ADAPT ensembles and the work presented here. The ADAPT model is driven by observed line-of-sight (LOS) or vector magnetograms incorporated into the existing model state using an ensemble least squares method \citep{Hickmann2015} over the observed solar disk. The state of the ADAPT model ensemble is also influenced by the initializing seed map into which observations are assimilated, but the impact of this initial state decreases over time as more observational data are assimilated. ADAPT can run with $0.2\degr$ spatial resolution in Carrington latitude and longitude, but the public maps typically generated on a daily basis and used in this study are generated with $1\degr$ resolution. In addition, the public maps include spatial smoothing to preserve the observational input resolution in the model longitude and latitude reference frame. The ADAPT maps used in this study are generated daily at 20 UT and can be found online at the National Solar Observatory (NSO) site: \url{https://gong.nso.edu/adapt/maps/}.

\subsection{NSO KPVT Magnetograms}
\label{sec:data:KPVT}
To create the global maps for CR1901, the magnetograms used to drive the ADAPT model come from the Kitt Peak Vacuum Telescope \citep[KPVT;][]{Livingston1976} using the NASA/NSO Spectromagnetograph \citep{Jones1992}. The KPVT magnetograms were generated daily (weather permitting) over the course of an hour from spatially resolved (with 1.14~arc~second resolution) long-slit spectral scans in the \ion{Fe}{1}~868.8~nm line \citep{Jones2000} and the field was remapped assuming purely radial magnetic fields. During this period, only the magnetogram on October 22\textsuperscript{nd} was unobserved and excluded from assimilation into ADAPT. The magnetograms used to drive the ADAPT maps in this study are provided by the NSO and can be found online at \url{https://nispdata.nso.edu/ftp/kpvt/daily/rawsyn}.

\subsection{NSO Coronal Hole Map}
\label{sec:data:Harvey}
To estimate the regions of open flux at the poles, we use the CH map for CR1901 from \citet{Harvey2002}. This single map was derived from the KPVT \ion{He}{1}~1083~nm images and KPVT magnetograms taken synoptically over the entire rotation. The CH regions were identified by hand through a combination of their appearance in the images (brightness, network contrast), magnetic features ($>75\%$ unipolar), and size (at least two supergranules). We recognize that all CH boundary detection methods include various sources of uncertainty \citep[see, e.g.,][]{Linker2021, Reiss2021}. In future work, we intend to utilize boundary uncertainty estimates, however for this study we use the CH regions as determined by \citet{Harvey2002} as the ground truth. This boundary is plotted on an ADAPT map from the start of CR1901 in Figure \ref{fig:ADAPT_original} panel (a). The CH boundary data can be obtained from the NSO online at \url{https://nispdata.nso.edu/ftp/kpvt/coronal_holes}.

\subsection{WIND in situ Measurements}
\label{sec:data:WIND}
The SW measurements on the Sun-Earth line were recorded by the WIND spacecraft \citep{Acuna1995} with the wind speed measured by the Solar Wind Experiment \citep[SWE;][]{Ogilvie1995} and the interplanetary magnetic field (IMF) measured by the Magnetic Field Investigation \citep[MFI;][]{Lepping1995}. At the time of these observations, WIND had not yet achieved its final orbit around the L1 Lagrange point. During the CR1901 period, it was between the Earth and L1, $7.6$--$11.4\times10^5$~km upstream of Earth, while $0.5$--$1\times10^5$~km below the ecliptic, and from $2.4\times10^5$~km ahead to $0.6\times10^5$~km behind the Earth-Sun line. These deviations from the Earth-Sun line are equivalent to its eventual halo orbit around L1 but at about half to three quarters the distance from the Earth. The hourly averaged WIND observed SW speed and IMF polarity data used in this study are distributed with WSA and can also be obtained online at \url{https://omniweb.gsfc.nasa.gov}.

\subsection{Ulysses in situ Measurements}
\label{sec:data:Ulysses}
The Ulysses spacecraft \citep{Wenzel1992} measured the SW with the Solar Wind Plasma Experiment \citep[SWOOPS;][]{Bame1992} and the IMF with the Magnetic Field Investigation \citep{Balogh1992} on a highly elliptical and inclined orbit specifically to sample the heliosphere out of the ecliptic. During the CR1901 period, Ulysses was executing a slow climb from $2.36$ to $2.48$~AU above the ecliptic after passing over the solar north pole, traversing $70$--$64.3\degr$ in heliographic latitude at a radial distance of $2.44$--$2.63$~AU leading the Earth by $123$--$102\degr$ in longitude. The hourly averaged Ulysses observed SW speed and IMF polarity data used in this study are distributed with WSA and can also be obtained online at \url{https://omniweb.gsfc.nasa.gov}.

\section{Modifying ADAPT Modeled Polar Flux}
\label{sec:ADAPT}
To date, individual solar magnetogram observations of the polar regions are generally too noisy to be assimilated into the ADAPT model ensemble. The polar observational uncertainties arise primarily from foreshortening and the canopy effect, combined with the highly variable horizontal magnetic signal that increases toward the limb \citep[e.g.;][]{Harvey2007}. To help minimize the injection of observational uncertainty into the global maps, the ADAPT model limits the ingestion of observed magnetic fields to low latitudes ($\sim\pm60\degr\pm7\degr$) depending on the input magnetogram and the solar $b$ angle. Consequently, the polar field distributions in the models are determined entirely by the advection of low-latitude fields towards the poles by (primarily) meridional circulation combined with the random emergence and motion of flux in the polar regions. This process yields average flux densities and evolution consistent with observations \citep[see Figure 11 of][]{Posner2021}, but the polar field distribution is unconstrained at any given time. This is particularly problematic because the prominent dipole field component, especially during solar minimum, is driven largely by the polar fields \citep{Wang2009}.

Despite this lack of polar field observations, it is possible to infer the general distribution of polar magnetic fields based on observed polar CHs. Polar CHs are known to contain highly unipolar magnetic fields \citep[more than 70\% of the area contains fields of the dominant polarity;][]{Harvey2002, Henney2005}, and the dominant polarity of each polar region (if not the pole itself) is reliably discernible \citep[e.g.,][]{Wang2009} even during the polar field reversals at solar maximum \citep[e.g.,][]{Babcock1959}. Exploiting these known characteristics, it is possible to constrain the magnetic fields near the poles of ADAPT maps in a statistical sense using observed polar CHs.

\subsection{Polar Flux Modification}
\label{sec:ADAPT:modification}
The basic premise of the flux modification procedure is to adjust the ADAPT modeled polar magnetic flux to be consistent with observed CH boundaries. This is accomplished by enforcing a minimum unipolarity (percentage of the area of a particular polarity) within and outside CH boundaries. In this work, the polar holes are modified to at least $90\%$ of the dominant polarity while outside the CHs are modified to at least $55\%$ of the opposite polarity. These highly unipolar CHs are actually less uniform than determined by \citet{Harvey2002} who found $98.3\%$ and $98.5\%$ unipolarity for the northern and southern polar CHs during the CR1901 period. However, the percent unipolarity depends on the magnetic map resolution and, while the \citet{Harvey2002} observations have $1\degr$ resolution in longitude, the sine-latitude mapping has larger area pixels near the poles that results in higher measured unipolarity. In addition, WSA is relatively insensitive to the precise percent unipolarity in the polar regions, so this discrepancy has only a minimal impact on the results. The polar field modification is applied only to the polar regions above a constant latitude cutoff, and the modification is applied independently for each member of the ADAPT ensemble on each day of the studied period.

We modify the polar fields by randomly shuffling fluxes in the polar regions to conform to the $90\%$--$55\%$ unipolarity constraints defined by the observed CH boundaries. This procedure completely removes the spatial flux distribution in the polar regions of the ADAPT model that is driven by the random-walk nature of supergranulation flows. The flux redistribution process is applied to the two poles independently, treating areas inside and outside the polar CHs separately, and conserves positive, negative, net, and total magnetic flux within each modified pole as a whole. A more detailed algorithmic description of this flux redistribution is given in Appendix \ref{sec:appendix:randomization_algorithm}. We also tested a method based on scaling the ADAPT polar fluxes that conserved the spatial structure of the polar fields but found that it did not perform as well as the flux randomization technique. This procedure and example results are described in Appendix \ref{sec:appendix:scaling_algorithm}.

\subsubsection{Defining the Low-Latitude Cutoff}
\label{sec:ADAPT:cutoff}
An important component of the flux modification procedure is the definition of the cutoff above which the polar fields are modified. The choice of this boundary latitude is a balance between the desire to significantly modify the polar regions (particularly to encompass the observed CH boundaries) while also respecting the well-observed low-latitude fields in ADAPT. We adopt two different choices of the boundaries. In the more conservative case, we modify all the flux outside the ADAPT data assimilation range which, during the CR1901 period spanned latitudes from $-56\degr$ to $+61\degr$. The more aggressive scenario modifies all flux above $\pm45\degr$ latitude to fully encompass the observed CH boundaries.

\subsubsection{Polar Magnetic Field Strength}
\label{sec:ADAPT:strength}
There is a long-standing discrepancy between interplanetary magnetic field (IMF) strength observed by in situ spacecraft vs. extrapolations from global magnetic maps \citep[e.g.,][and references therein]{Linker2017}. In general, extrapolations of the photospheric magnetic field are found to underestimate the observed IMF by up to factors of a few. A detailed study by \citet{Wang2022} suggests that accounting for saturation of the observed magnetically sensitive lines can significantly reduce this discrepancy for some magnetographs, but incorporating high latitude field observations into global magnetic maps also poses particular difficulties. Sources of polar magnetic field strength uncertainty include, e.g.: 1) data assimilation misalignment with latitude between old and new observations; 2) how the polar observational gaps are filled and estimated; 3) foreshortening and the highly variable horizontal magnetic signal that increases toward the limb \citep[e.g.,][]{Harvey2007}. Each source of data assimilation uncertainty can lead to flux cancellation and/or weakening during each data assimilation step near the poles. We therefore also investigate the impact of the polar magnetic field strength by performing very preliminary tests of simply doubling the magnetic field strengths at the poles, using the more conservative low-latitude cutoff.

\subsection{Example Polar Flux Modification}
\label{sec:ADAPT:effects}
\begin{figure*}[!ht]
    \includegraphics[width=\textwidth]{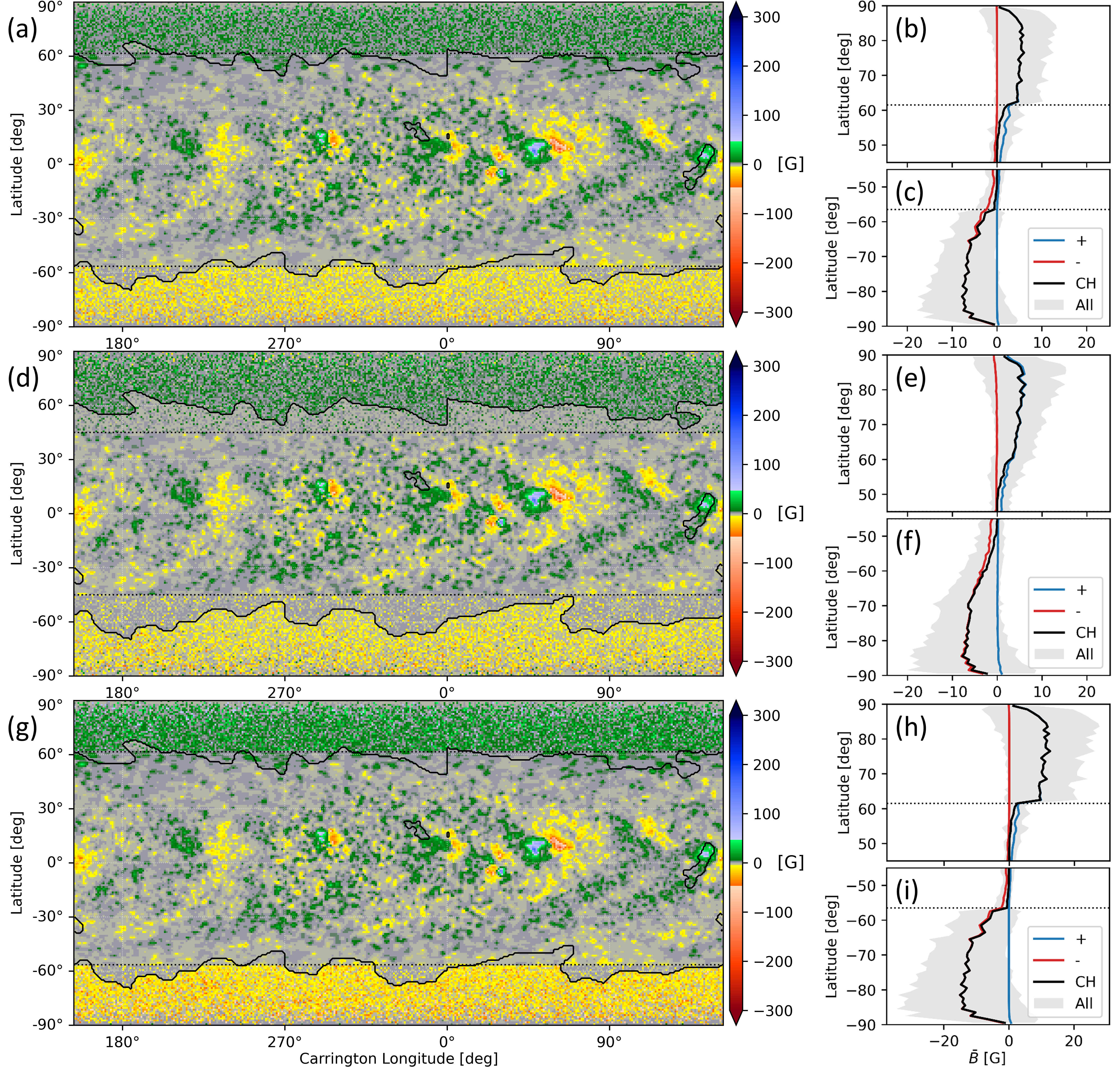}
    \caption{Same as Figure \ref{fig:ADAPT_original} except for the ADAPT ensembles with: modified polar flux (panels (a), (b), and (c)); modified polar flux down to $\pm45\degr$ latitude (panels (d), (e), and (f)); and polar flux modified and doubled (panels (g), (h), and (i)). In all panels, the horizontal dotted black lines indicate the latitude boundaries between which the flux is unmodified. In panels (d), (e), and (f), the dotted lines are at $\pm45\degr$ while in all other panels they indicate the ADAPT data assimilation range at $-56\degr$ and $+61\degr$.
    \label{fig:ADAPT_modified}}
\end{figure*}

Characteristic examples of the randomized flux modification are shown in Figure \ref{fig:ADAPT_modified} for the first realization of the respective ADAPT ensembles on 1995 October 1\textsuperscript{st}. The effect of the randomization is obvious, resulting in much ``noisier,'' more uniform spatial distributions of the polar fields. The randomization process yields sharp edges in the magnetic field distribution along the CH and low-latitude cutoffs compared to the original ADAPT map in Figure \ref{fig:ADAPT_original} in which the field distribution appears only loosely related to the observed CH boundaries. Note that the $-56\degr$ and $+61\degr$ cutoffs do not fully encompass the observed CH boundaries. This is particularly true in the northern hemisphere which is tilted towards the Earth at this time, allowing ADAPT to assimilate observations up to higher latitudes. In this example, the high-latitude polar regions are nearly completely (more than $90\%$) unipolar, positive in the north and negative in the south. Note also that in the modified regions outside of the polar CHs (that have the opposite polarity over $55\%$ of the area by construction) the polarity of the net flux is the same as the CH region because the opposite polarity field strengths are typically weak.

Comparing panels (b) and (c) in Figure \ref{fig:ADAPT_original} with (b), (c), (e), and (f) in Figure \ref{fig:ADAPT_modified}, it is clear that the flux randomization introduces only small changes in the latitudinal flux profiles. The randomization increases the flux density variability at a given latitude, shifts the latitude of peak flux density from $\sim75\degr$ to $\sim85\degr$, and flattens the distributions in general, reducing the maximum flux density but increasing it somewhat near the poles and the low-latitude boundary. However, the small amplitude of these changes relative to the standard deviation as a function of latitude (represented by the gray shaded region) suggests that these shifts in the latitudinal flux density profile should have only minor impacts on resulting coronal magnetic field models. The latitude profiles of these three ADAPT maps are roughly consistent with the range in Hinode \citep{Kosugi2007} Solar Optical Telescope \citep{Tsuneta2008} Spectro-Polarimeter \citep{Lites2013} polar observations \citep[Figures 8 and 9 of][]{Petrie2017}. The doubled polar field strengths in panel (g) are immediately evident in the associated profiles in panels (h) and (i). These average field strengths are significantly larger than indicated by the \citet{Petrie2017} observations.

\section{WSA models}
\label{sec:WSA}
The Wang-Sheeley-Arge (WSA) SW model \citep{Arge2000, Arge2003, Arge2004a} is a data-driven model of the solar corona, capable of predicting the in situ SW speed and magnetic field polarity anywhere in the heliosphere using input from global photospheric magnetic field maps. WSA combines a global coronal magnetic field solution (out to a user-defined outer radius $R_o$, typically 5 or 21.5~$R_\odot$) with a radially propagating and interacting SW evolving purely radially, creating a characteristic Parker spiral \citep{Parker1958}. For this study we use version 5.3.2 of the WSA model, using the key WSA model run parameter values listed in Appendix \ref{sec:appendix:parameters}. One particularly relevant change to this version of WSA relating to the inner boundary radius and how field line tracing is used to determine the modeled open field is described in Appendix \ref{sec:appendix:inner_boundary}.

The WSA coronal magnetic field is calculated through a combination of a potential field source surface \citep[PFSS;][]{Schatten1969, Altschuler1969, Wang1992} magnetic field extrapolation from the photosphere out to the source surface radius ($R_{ss}$) and the Schatten current sheet \citep[SCS;][]{Schatten1971} model from the interface radius ($R_i$) out to the coronal boundary at $R_o$. By definition, $R_\odot<R_i<R_{ss}<R_o$, and the choice of $R_i$ and $R_{ss}$ significantly influences the fidelity of the global magnetic field solution \citep{McGregor2008, Lee2011, Arden2014, Virtanen2020}. \citet{Meadors2020} investigated the coronal solutions during CR1901 and found optimal radii ($R_{ss}$,~$R_i$) = (3.5-3.9~$R_\odot$, 3.0-3.4~$R_\odot$) with both radii evolving towards the smaller values over the course of the rotation. These are significantly larger than the canonical ($R_{ss}$,~$R_i$) = (2.51~$R_\odot$, 2.49~$R_\odot$) radii \citep{Altschuler1969, Arge2003} and result in smoother, more realistic field solutions at $R_i$ and SW predictions at WIND that are much more consistent with the observations. In addition, they find that the footpoints of the magnetic connectivity to WIND shift to higher latitudes, deeper into the polar CHs. For comparison, we compute both ``standard solutions'' ($R_{ss}$,~$R_i$) = (2.51~$R_\odot$, 2.49~$R_\odot$) and optimized solutions ($R_{ss}$,~$R_i$) = (3.7~$R_\odot$, 3.2~$R_\odot$) with $R_o=5R_\odot$.

The SW speed ($v_{sw}$) at $R_o$ is calculated from the magnetic field solution using a combination of the flux tube expansion factor $f_{ss}$ \citep{Wang1990} and the angular distance to the nearest CH boundary $\theta_b$ \citep{Riley2001} according to
\begin{equation}
\label{eqn:WSA:speed}
v_{sw}\left(f_{ss},\theta_b\right) = v_0 + \left(\!\frac{v_m}{\left(1+f_{ss}\right)^{c_1}}\!\right)\!\!\left(\!1-c_2\exp\!\left(\!-\!\left(\!\frac{\theta_b}{c_3}\!\right)^{\!\!c_4}\right)\!\!\right)^{\!\!c_5}
\end{equation}
where $v_0=285$~km~s\textsuperscript{-1} is the minimum velocity, $v_0+v_m=910$~km~s\textsuperscript{-1} is the maximum velocity, and $c_1=\frac{2}{9}$, $c_2=0.8$, $c_3=2.0$, $c_4=2.0$, and $c_5=3.0$ are magnetic map and model specific empirical constants. The flux tube expansion factor is calculated for each point on the outer boundary as
\begin{equation}
\label{eqn:WSA:fs}
f_{ss}=\left(\frac{R_{\odot}}{R_{ss}}\right)^{\!\!2}\!\!\left(\frac{B_{\odot}}{B_{ss}}\right)
\end{equation}
where $B_{\odot}$ and $B_{ss}$ are the magnetic field magnitude on the photosphere and source surface connected by the magnetic model. From the outer boundary, the SW propagates outward in $\sim$1/10\textsuperscript{th} AU steps, with radially adjacent cells interacting such that faster moving parcels are slowed so as not to overtake slower parcels ahead of them \citep[for details see][]{Arge2000}. By specifying the path of a (natural or artificial) satellite in the heliosphere, WSA predicts the associated SW speed and magnetic field polarity at that satellite which can be compared with observations.

\subsection{Predicted Coronal Holes}
\label{sec:WSA:coronal_hole}
\begin{figure*}[!t]
    \includegraphics[width=\textwidth]{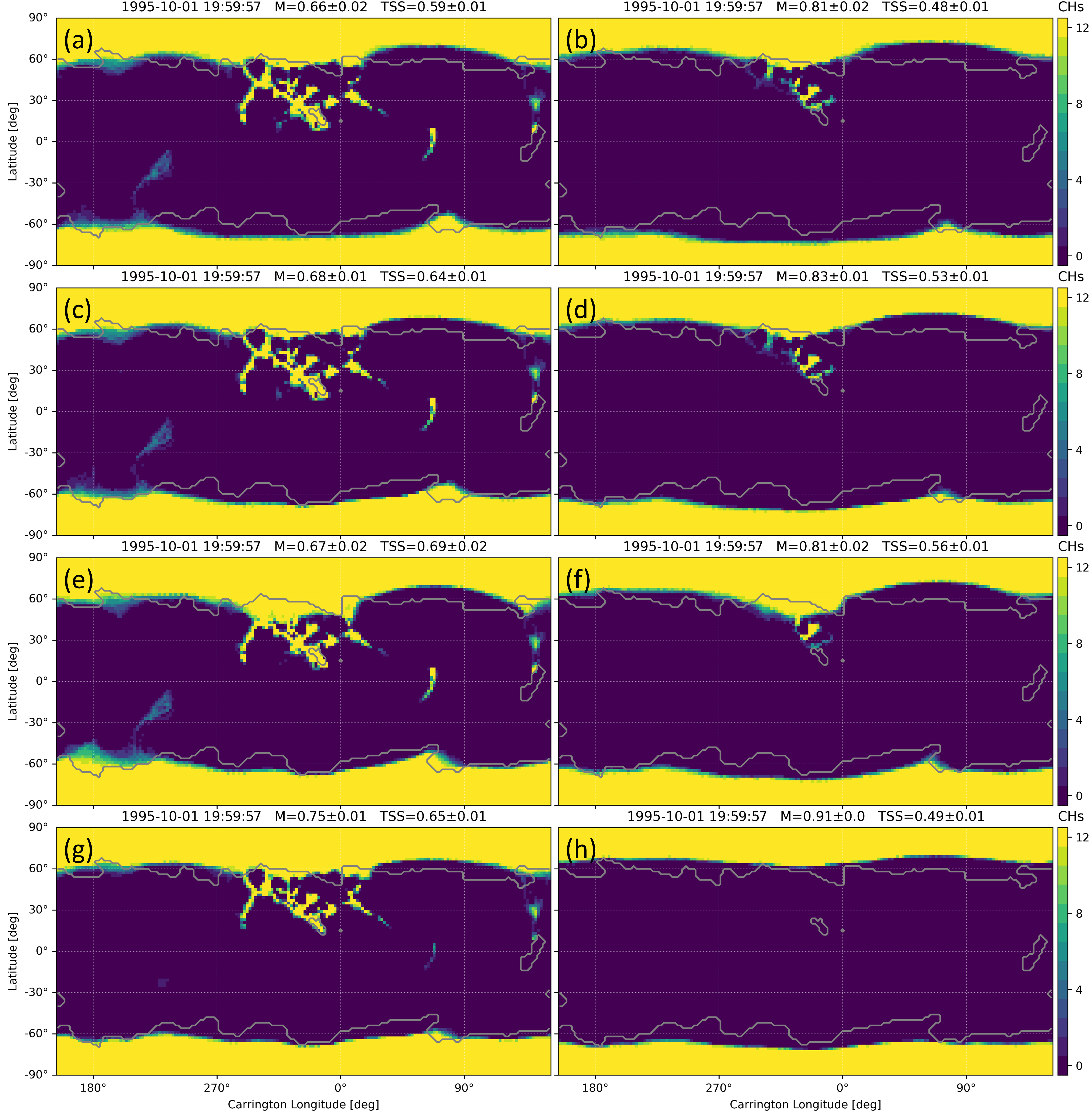}
    \caption{Observed and modeled coronal holes (CHs) on 1995 October 1\textsuperscript{st} generated with WSA driven by the four ADAPT ensembles. Panels (a) and (b) show the original, unmodified ADAPT ensemble; panels (c) and (d) show the ensemble with modified polar flux; panels (e) and (f) show the ensemble with the polar flux modified down to $\pm45\degr$; and panels (g) and (h) show the ensemble with polar flux modified and doubled. In panels (a), (c), (e), and (g) WSA solved for the coronal magnetic field using ($R_{ss}$,~$R_i$) = (2.51~$R_\odot$, 2.49~$R_\odot$) while in panels (b), (d), (f), and (h) WSA used ($R_{ss}$,~$R_i$) = (3.7~$R_\odot$, 3.2~$R_\odot$). In each panel, the color image indicates the number of realizations in the ensemble in which WSA identified the pixel as a CH. The gray contours indicate the observed \protect\citet{Harvey2002} CH boundaries. The title of each panel also indicates the Markedness (M) and true skill statistic (TSS) of the WSA ensemble prediction.
    \label{fig:CHs}}
\end{figure*}

As part of the WSA processing pipeline, the open photospheric magnetic fields are determined from the combined PFSS and SCS magnetic model components, allowing for direct comparisons between observed CHs and the modeled open fields estimated by WSA. If we utilize observations as the ground truth, then the WSA model is attempting to ``classify'' if each pixel in the open-field map is part of an observed CH. In that case, each pixel in a map is a test, with the WSA modeled map either correctly or incorrectly (True or False) identifying a CH or non-CH (Positive or Negative). We can then consider the Markedness (M) and true skill statistic (TSS, also known as the Informedness or the Youden J statistic \citep{Youden1950}) of these predicted CH maps
\begin{equation}
\begin{aligned}
    \label{eqn:markedness_informedness}
    \text{M} &=\frac{TP}{TP+FP}-\frac{FN}{FN+TN} \\[5pt]
    \text{TSS} &=\frac{TP}{TP+FN}-\frac{FP}{FP+TN}
\end{aligned}
\end{equation}
that each measure the performance of the prediction from 0 (chance) to 1 \citep[perfect;][]{Hanssen1965, Powers2011, Chicco2021}. These metrics are complimentary, with M measuring the ability of the system to make correct predictions and TSS indicating how much of reality the prediction captures. Crucially, each considers true negatives, correctly rewarding the correspondence between predicted and observed non-CH regions. This is an important consideration when evaluating a predictor with imbalanced data \citep{Woodcock1976, Bloomfield2012}, when the real number of positives and negatives is unequal, as is generally the case with the CH and non-CH areas. Calculation of the M and TSS is performed on the areas (not the number of pixels) in the observed and predicted CH maps because of the dependence of the pixel areas on latitude. For example, TP represents not the number of pixels but the fractional area of the solar surface where CHs are correctly identified. Note also that these metrics are calculated over the entire maps (which include isolated low-latitude CHs) not just the polar regions.

\begin{figure*}[!t]
    \includegraphics[width=\textwidth]{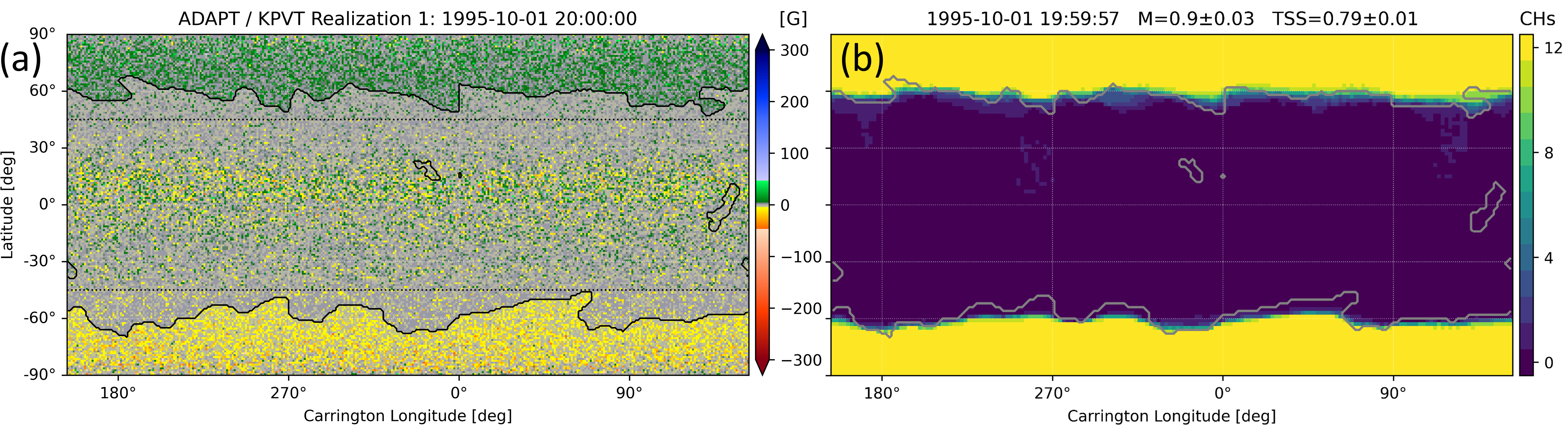}
    \caption{The ADAPT ensemble with the polar flux modified down to $\pm45\degr$ and additional low-latitude field shuffling to remove active region field concentrations on 1995 October 1\textsuperscript{st}. Panel (a) shows the radial photospheric magnetic field of the first ADAPT realization with horizontal dotted black lines at $\pm45\degr$ (similar to panel (d) in Figure \ref{fig:ADAPT_modified}). Panel (b) shows the modeled coronal holes from WSA with the coronal magnetic field solved using ($R_{ss}$,~$R_i$) = (2.51~$R_\odot$, 2.49~$R_\odot$) with the Markedness (M) and true skill statistic (TSS) of the ensemble prediction indicated in the title (similar to panel (e) in Figure~\ref{fig:CHs}).
    \label{fig:CHs_shuffle}}
\end{figure*}

Using the different WSA radii and polar region processing, eight ensemble CH solutions and their associated metrics (with standard deviations) for October 1\textsuperscript{st} are shown in Figure~\ref{fig:CHs}. All of the maps have large polar CHs that are broadly consistent with the observation but poorly capture the observed low-latitude CHs. From these maps it is clear that changing the PFSS and SCS solution radii (left/right) has a much larger impact on the CH areas than modifying the polar flux. This is particularly obvious when considering the low-latitude CHs which nearly completely disappear when using the larger radii. The larger solution radii result in increased M and decreased TSS. That is, a positive or negative prediction of a CH is more likely to be correct, but less of the observed CH area is correctly identified by the model. These changes are the result of the process by which the PFSS and SCS models are combined. At $R_i$, all open fields in the PFSS model are considered open in the SCS model and therefore define the WSA CHs. By increasing $R_i$, more magnetic field is allowed to close in the PFSS model (independent of $R_{ss}$), therefore reducing the open field area on the photosphere and resulting in smaller CHs. In this case, most (or all in the doubled field strength model in panel (h)) of the low-latitude CHs actually appear closed by R=3.2$R_\odot$. 

On the other hand, for each set of solution radii, applying the polar flux redistribution typically improves the metrics somewhat and never worsens them. In general, the predicted boundaries approach the observed boundaries and exhibit smaller variation between realizations, but these effects require close inspection to identify. Unsurprisingly, the largest changes occur for the scenarios with double the polar field strengths (panels (g) and (h), in Figure~\ref{fig:CHs}) which cause a dramatic reduction in the total predicted CH areas. In the scenario with the smaller WSA solution radii (panel (g)), this results in the best agreement between the models and observations (particularly in the southern hemisphere near $180\degr$E), whereas the larger WSA solution radii (panel (h)) cause the modeled boundaries to recede from the observed boundaries in some places (e.g. that same location in the southern hemisphere near $180\degr$E). The scenarios with the more aggressive low-latitude cutoffs (panels (e) and (f)) in general cause the polar CHs to expand towards lower latitudes. Where the modeled CH boundaries from the unmodified maps are at higher latitudes than the observations this yields improvements, but the low-latitude extension at $50\degr$N $315\degr$E and the region near $-60\degr$S $180\degr$E both expand well past the observed CH boundaries. The ``standard'' polar flux modification using the more conservative low-latitude cutoff and the unmodified polar field strengths (panels (c) and (d)) yields the most subtle improvements, but consistently shifts the modeled boundaries towards the observations without significant over correction like the other two redistribution scenarios.

While improved by the various flux modifications, in no scenario do the modeled ensemble polar CH boundaries closely resemble the observed boundaries. This is particularly true for those tests where all the fields above $\pm45\degr$ latitude are modified, fully capturing the observed polar CH boundaries, but the modeled boundaries maintain the same large-scale structure as those from the unmodified ADAPT maps. Combined with the poor modeling of the isolated low-latitude CHs, this suggests that the large-scale structure of the WSA modeled polar CHs depends strongly on the low-latitude magnetic fields that remain unmodified in all the ADAPT ensembles for the CR1901 period maps.

To test the importance of active regions in defining the polar CH boundaries we remove the active regions from all the ADAPT maps in one ensemble and use it to drive WSA (with the smaller solution radii). We start with the polar modified ensemble with the $\pm45\degr$ latitude cutoffs (to ensure that the polar field fully captures the observed CH boundaries) and then shuffle all flux between $\pm45\degr$ latitude in longitude, i.e. the longitudinal position of each flux element at a given latitude is randomized. An example ADAPT map after the lower latitude flux reshuffling and the resulting WSA CH map are shown in Figure~\ref{fig:CHs_shuffle}. The modeled polar CHs now closely conform to the observed boundaries and the metrics are significantly improved. This also completely destroys the modeled low-latitude CHs which were in any case not well captured in the previous models.

Obviously, shuffling the low-latitude flux randomly is not meant to realistically represent the photospheric magnetic field. Rather, this demonstrates that the low-latitude field distribution in the ADAPT ensemble during the CR1901 period does not accurately estimate the distribution of fields on the Sun. The lower latitude issues in the ADAPT maps could be due to problems with the flux-transport-component in ADAPT, inherent limitations with the remapping and data assimilation, or missing flux emergence from new or evolving active regions on the farside. Additional difficulties can arise if only part of an active region rotates into the data assimilation window as that ADAPT map will then attempt to assimilate significantly imbalanced flux which can cause changes that ripple across the whole map. Whatever the cause, inaccurate or incomplete knowledge of active region magnetic fields appears to significantly impact polar CH boundaries, even when the polar flux modification is applied.

\subsection{Low-latitude WIND prediction}
\label{sec:WSA:WIND}
\begin{figure*}[!t]
    \includegraphics[width=\textwidth]{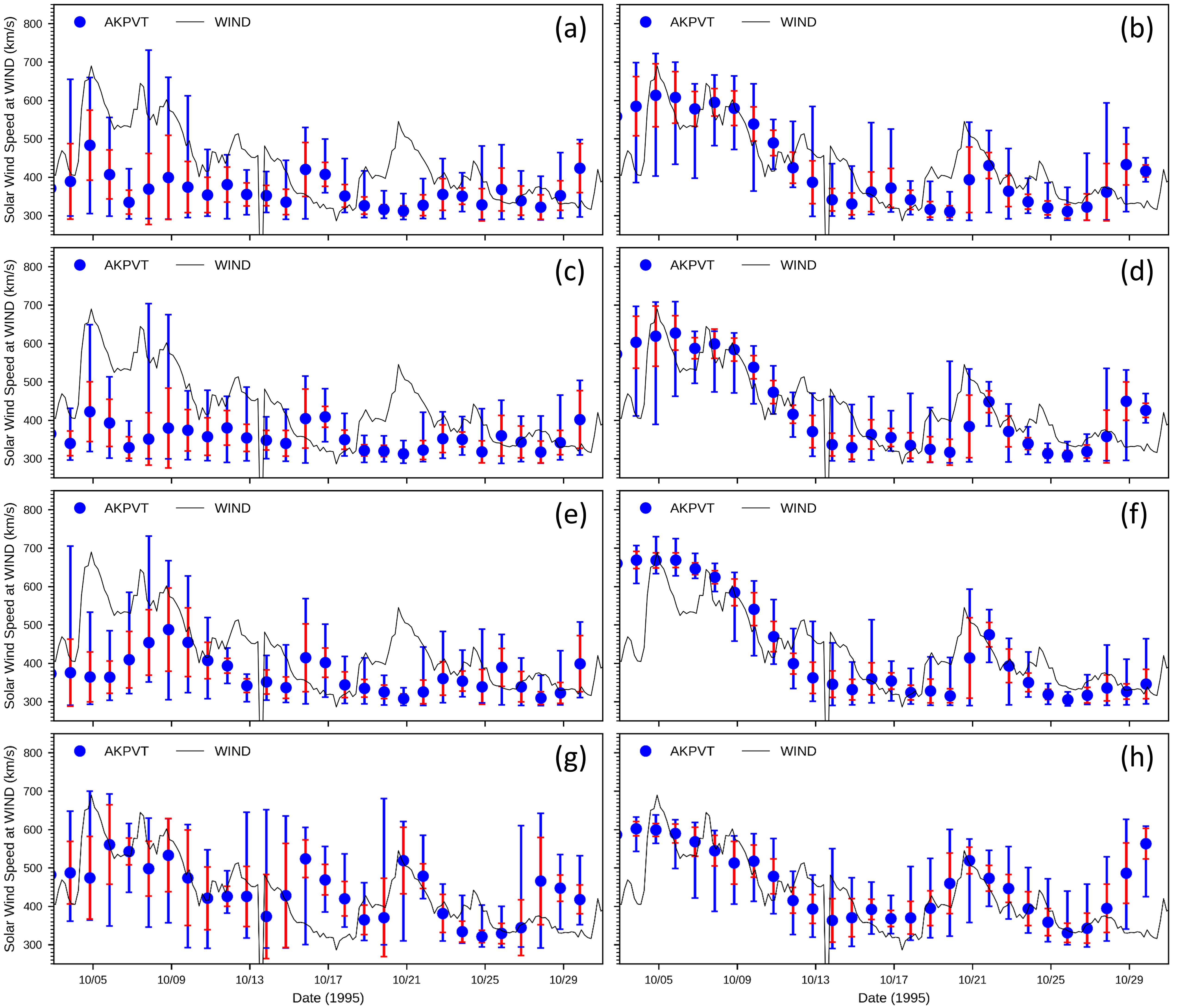}
    \caption{The WSA 3-day solar wind speed predictions at the WIND spacecraft over CR1901 driven by the four ADAPT ensembles compared with observations. The panel order is the same as in Figure \ref{fig:CHs}. In each panel, the black line indicates the observations, the blue points indicate the predicted ensemble mean at 20 UT, and the blue (red) bars indicate the ensemble range (standard deviation).
    \label{fig:WIND:timeseries}}
\end{figure*}

Sampling in the ecliptic near the Sun-Earth line, WIND measures the SW emitted from relatively low solar latitudes, typically from low-latitude CHs, active region open fields, or near the boundary of polar CHs. We expect, therefore, that the predictions at WIND will be particularly sensitive to the details of the CH boundary. Additionally, the diversity of the solar sources of low-heliospheric latitude wind, even during solar minimum, means that the wind speed typically varies greatly over a single solar rotation \citep{McComas1998a}. We see in Figure \ref{fig:WIND:timeseries} that this is the case during CR1901 with the SW speed observed at 300--700~km~s\textsuperscript{-1}. We investigate the 3-day WSA predictions at WIND since that is the characteristic travel time for the SW from the Sun to the Earth (i.e. $\sim580$~km~s\textsuperscript{-1} wind).

Figure \ref{fig:WIND:timeseries} illustrates that the standard ADAPT ensemble and WSA solutions with ($R_{ss}$,~$R_i$) = (2.51~$R_\odot$, 2.49~$R_\odot$) (panel (a)) does not well characterize the SW speed observed at WIND during CR1901. The ensemble mean does not predict the higher speed wind during October 4\textsuperscript{th}--14\textsuperscript{th} (although some individual realizations do capture parts of this observed speed increase) and 18\textsuperscript{th}--21\textsuperscript{st} and identifies a spurious increase to $\sim400$~km~s\textsuperscript{-1} on October 15\textsuperscript{th} and 16\textsuperscript{th}. The polar flux modifications improve the predictions somewhat and, due to the homogenization of the polar magnetic fields, typically reduce the spread between ensemble members on a given day, but they still do not match observed values for the periods of high-speed wind and over-predict the speed between them. The exception to this is the prediction generated from the modified maps with double the polar field strength (panel (g) of Figure \ref{fig:WIND:timeseries}) which captures both high speed streams well while over-predicting the wind speed during October 15\textsuperscript{th}--17\textsuperscript{th} (even more than the other maps investigated here) and October 27\textsuperscript{th}--29\textsuperscript{th}.

The predictions improve dramatically when the WSA coronal field is generated using ($R_{ss}$,~$R_i$) = (3.7~$R_\odot$, 3.2~$R_\odot$) (panel (b) of Figure \ref{fig:WIND:timeseries}), consistent with the improvements noted by \citet{Meadors2020} when using larger solution radii during this period. In particular, the higher speed wind during October 4\textsuperscript{th}--11\textsuperscript{th} is well characterized (including some of the substructure) and the prediction more closely captures the increased speeds on October 20\textsuperscript{th} and 21\textsuperscript{st}. Even more than for the models with the smaller solution radii, the polar flux modification typically decreases the spread in the ensemble. The exceptions are the predictions of slow wind from the models driven with the doubled polar field strength for which the ensemble spread tends to increase. This model is also notable as the only one that fully captures the speed during the 18\textsuperscript{th}--27\textsuperscript{th}, although it then over predicts the speed on the 28\textsuperscript{th} and 29\textsuperscript{th}. The ensemble averages of these models with the larger solution radii are unable to capture the increased speed on the 12\textsuperscript{th}--14\textsuperscript{th} and they all over predict the speed on the 2\textsuperscript{nd} and 3\textsuperscript{rd}. Despite these small discrepancies, the predictions from the models using the larger WSA solution radii perform significantly better, while the various flux modifications decreases the spread between individual ensemble members.

\begin{figure*}[!t]
    \includegraphics[width=\textwidth]{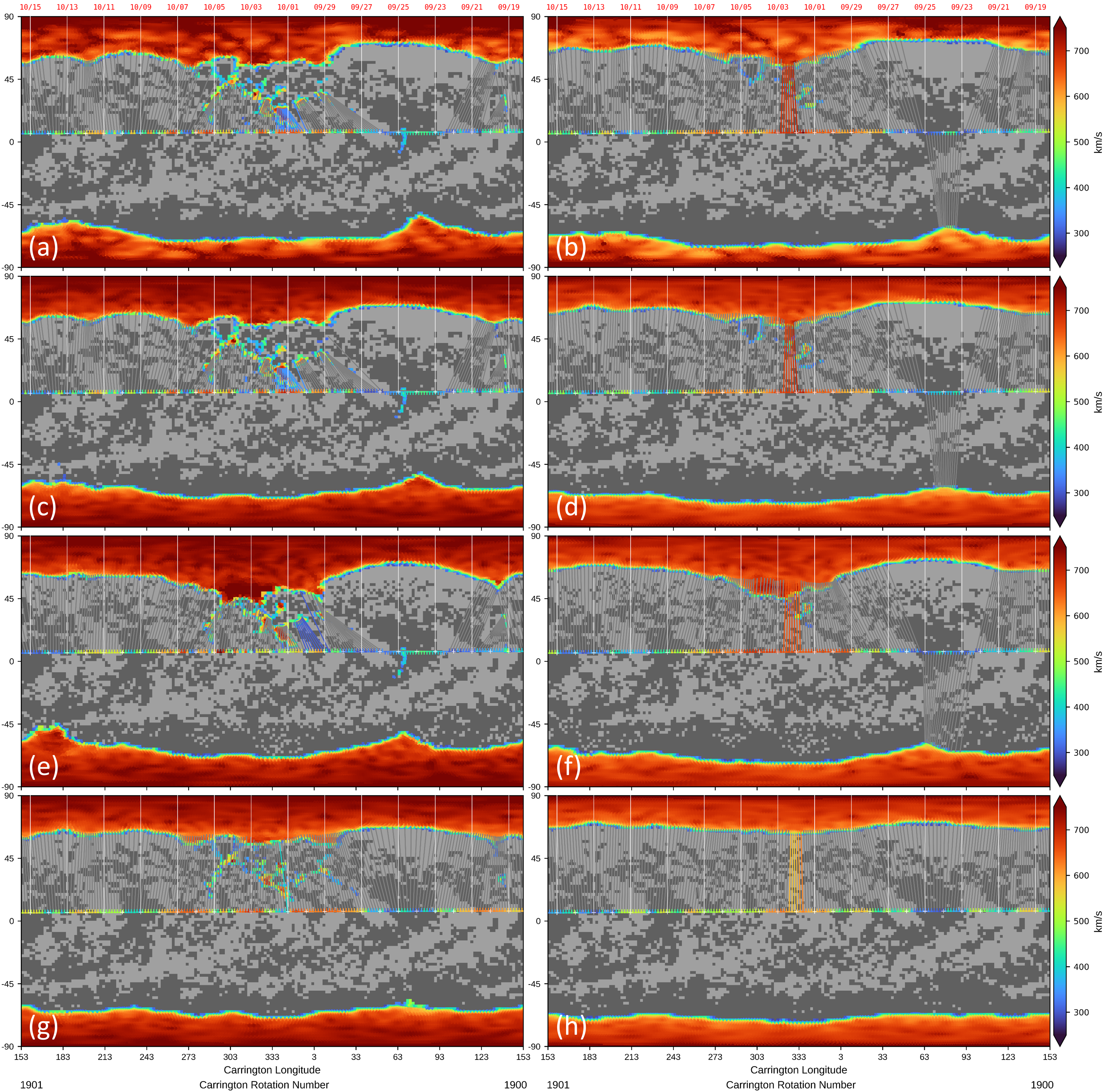}
    \caption{The WSA predicted coronal holes (CHs), solar wind (SW) speed, and WIND spacecraft connectivity from the first realization of the ADAPT ensemble on 1995 October 1\textsuperscript{st} for various polar magnetic field modifications. The panel order is the same as in Figures \ref{fig:CHs} and \ref{fig:WIND:timeseries}. The gray scale background image indicates the magnetic field polarity of the input ADAPT maps, where light gray is positive and dark gray is negative. The colored points indicate the predicted SW speed (calculated using equation \ref{eqn:WSA:speed}) from the photospheric foot points of open magnetic fields. The horizontal line of points near the equator indicates the sub-satellite track during this Carrington rotation, with the color indicating the SW velocity predicted at the outer boundary of the coronal model and the point direction indicating the positive ($^\bot$) or negative ($_\top$) magnetic field polarity. The (mostly gray) lines connecting these points to the open field regions indicate the mapping through the WSA coronal magnetic field model from the sub-satellite points back to the photosphere. The colored lines indicate connections where the wind predicted three days in advance at the spacecraft actually originates from this particular background magnetic field map, with the color indicating the predicted speed including propagation effects. The vertical white lines connecting the sub-satellite track and the labels above the figures indicate the dates at which the satellite was above that point on the Sun.
    \label{fig:WIND:derch}}
\end{figure*}

Figure \ref{fig:WIND:derch} illustrates why the WIND speed predictions improve when WSA is solved using the larger solution radii, particularly during the first high speed stream. In the models generated with the standard solution radii (excluding the models from the maps with double the polar field strength), the wind during the prediction window around October 4\textsuperscript{th} and 5\textsuperscript{th} (colored connecting lines) originates from small, low-latitude CHs resulting in generally lower speeds. Even with the polar flux modification, the changes in the CH boundaries have virtually no impact on the predicted wind speeds. Those models generated with the larger solution radii yield smaller low-latitude CHs leading to more connectivity to the northern polar CH. The same is true for the model generated using the smaller radii and double the polar field strengths which also exhibits smaller low-latitude CHs and has wind connectivity directly to the pole.

\subsection{High latitude Ulysses prediction}
\label{sec:WSA:Ulysses}
\begin{figure*}[!t]
    \includegraphics[width=\textwidth]{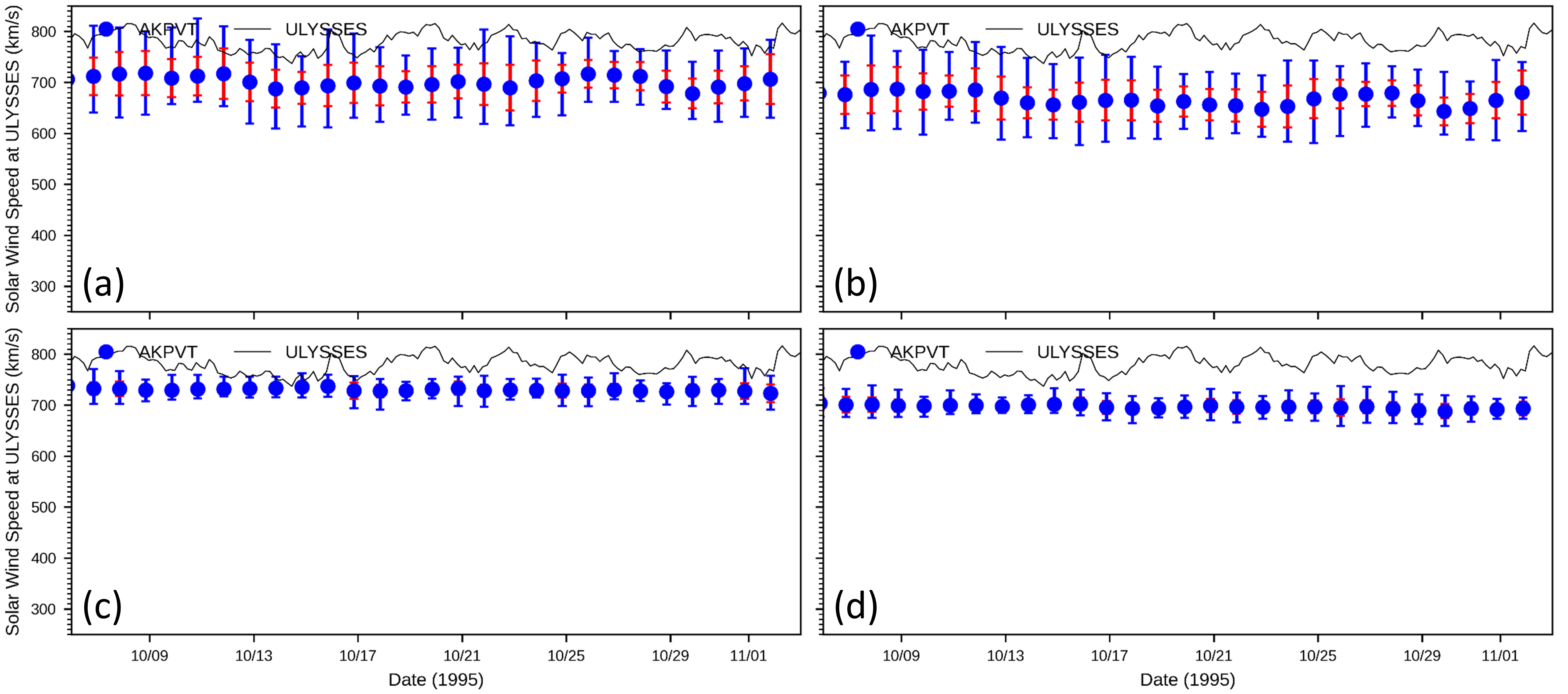}
    \caption{Same as Figure \ref{fig:WIND:timeseries} except for 6-day solar wind speed predictions at the Ulysses spacecraft compared with observations. Only panels (a), (b), (c), and (d) are shown as the results from the other polar modifications are nearly identical to those in panels (c) and (d).
    \label{fig:Ulysses:timeseries}}
\end{figure*}

The Ulysses spacecraft's position at high solar latitude during CR1901, coupled with the relatively simple magnetic field topology of solar minimum, results in the Ulysses spacecraft being connected to magnetic fields embedded deep within the northern polar CH. As illustrated in Figure \ref{fig:Ulysses:timeseries}, during this period Ulysses observed stable SW speeds of $780\pm17$~km~s\textsuperscript{-1} throughout the rotation. At these speeds, the SW reaches Ulysses in $\sim5.5$~days, so we evaluate the performance of the WSA 6-day predictions.

The WSA models driven with the unmodified ADAPT maps yield similarly stable predictions (compared with the observations) but with average speeds of $\sim700$~km~s\textsuperscript{-1} and $\sim665$~km~s\textsuperscript{-1} respectively for the smaller (panel (a)) and larger (panel (b)) solution radii. On a given day, the ensemble standard deviation characterizes the approximate spread in the observations over the entire rotation, with the outlying realizations typically differing by $\gtrsim100$~km~s\textsuperscript{-1}. The models driven with the flux-modified ADAPT maps all yield slightly increased average SW speeds (over the unmodified maps), $\sim725$--$730$~km~s\textsuperscript{-1} for the smaller solution radii and $\sim690$--$695$~km~s\textsuperscript{-1} for the larger, and the ensemble spreads are reduced dramatically. Of the three ADAPT modifications, the maps with double the polar field strength yield the slowest predicted speeds because of their associated smaller CHs, as discussed below. These increased wind speed predictions from the modified ADAPT maps are in better agreement with the observed SW, but they are well within the ensemble spread from the unmodified maps. In fact, because of that large spread, at almost all times there is at least one realization from the unmodified ensembles that performs better than the best realization from the modified ensemble. However, since these fluctuations are essentially random, each realization fluctuates about the ensemble mean, so there is no single realization from the unmodified maps that outperforms the modified map ensembles.

\begin{figure*}[!t]
    \includegraphics[width=\textwidth]{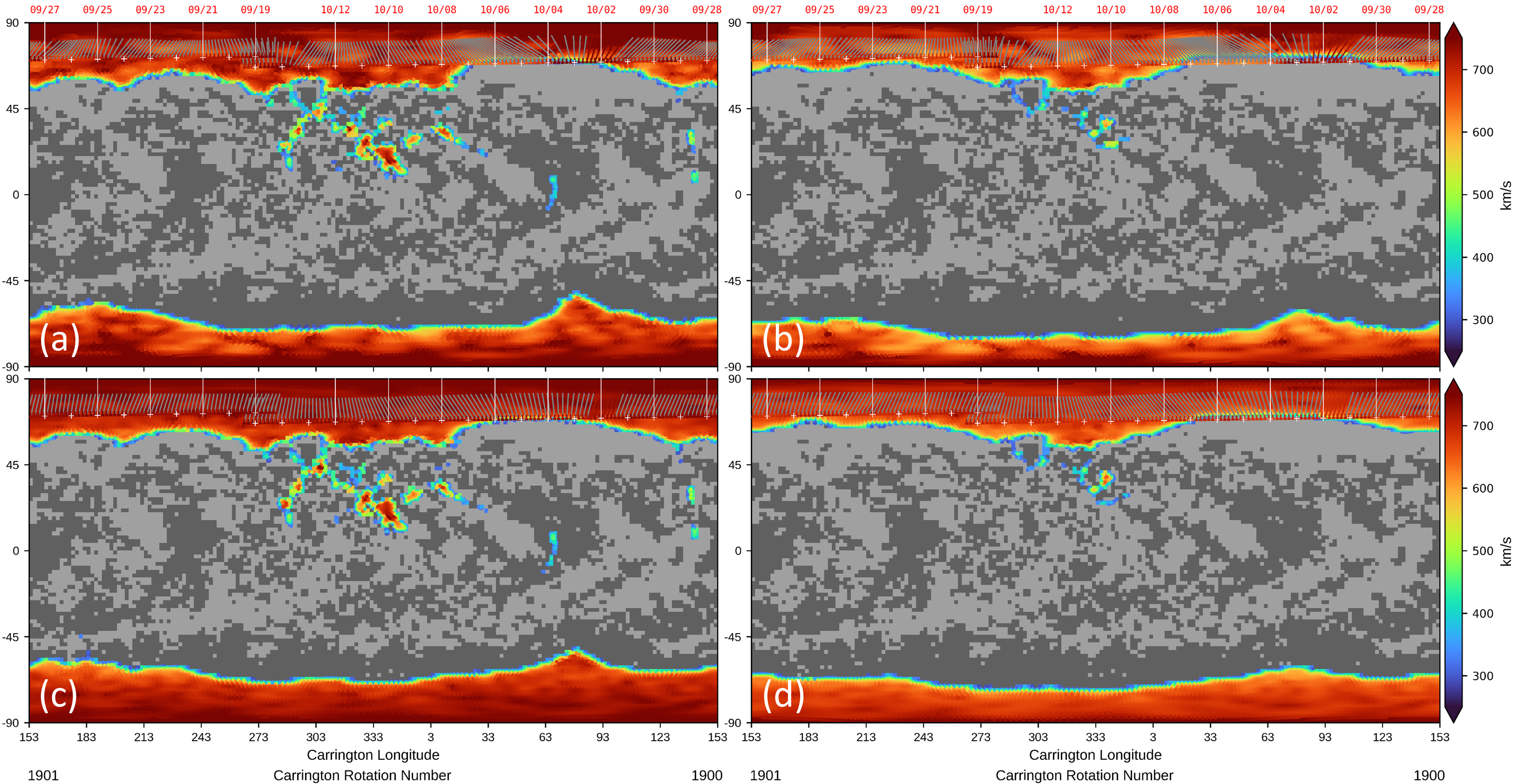}
    \caption{Same as Figure \ref{fig:WIND:derch} except for the 6-day connectivity and solar wind (SW) speed predictions at Ulysses. Only panels (a), (b), (c), and (d) are shown as the connectivity from the other polar modifications is nearly identical to those in panels (c) and (d). The photospheric magnetic polarity maps and predicted SW speed from the coronal holes are identical to those in Figure \ref{fig:WIND:derch}.
    \label{fig:Ulysses:derch}}
\end{figure*}

Consistent with our expectations, Figure \ref{fig:Ulysses:derch} confirms that during this period Ulysses was connected to wind originating from deep within the northern polar CH. As such, the changes to the polar CH boundaries from the various flux modifications have little impact on the predicted SW speed. However, by removing the structure of the polar fields, the flux modification procedure reduces the spatial variability of the wind speed predicted from the CH interior (as seen in the more uniform CHs in panels (c) and (d)) resulting in the much smaller spread in the ensemble predictions. We can also see that the reduced speeds predicted by the models using the larger solution radii are distributed mostly evenly across the polar CHs predicted from both the modified and unmodified ADAPT maps. The decreased wind speed predictions are due to the smaller polar CHs which require the field to, in general, expand more to fill the heliosphere. Because of the inverse relationship between the expansion factor and the predicted SW speed, this increased expansion results in slower predicted wind. We discuss this effect further in section \ref{sec:WSA:solar_wind}.

\subsection{The WSA prediction metric}
\label{sec:WSA:metrics}
\begin{table*}[!t]
\begin{threeparttable}
\caption{WSA Model Performance Metrics}
\centering
\begin{tabular}{l c c c c c c}
\hline
 ADAPT ensemble & $R_{ss}$ & $R_i$ & Markedness & TSS & \textit{H}(WIND)\textsuperscript{a} & \textit{H}(Ulysses)\textsuperscript{a} \\
\hline
  Original ADAPT  & 2.51 & 2.49 & 0.68$\pm$0.03 & 0.61$\pm$0.01 & 0.006$\pm$0.001 & 0.012$\pm$0.002 \\
  Modified poles  & 2.51 & 2.49 & 0.71$\pm$0.03 & \textbf{0.66$\pm$0.01} & 0.006$\pm$0.001 & \textbf{0.019$\pm$0.004} \\
  Modified, $\pm45\degr$  & 2.51 & 2.49 & 0.69$\pm$0.03 & \textbf{0.70$\pm$0.01} & 0.007$\pm$0.002 & \textbf{0.019$\pm$0.004} \\
  Modified, field $\times$2 & 2.51 & 2.49 & \textbf{0.79$\pm$0.03} & \textbf{0.66$\pm$0.01} & \textbf{0.009$\pm$0.004} & \textbf{0.017$\pm$0.003} \\
\hline
  Original ADAPT  & 3.7 & 3.2 & 0.84$\pm$0.04 & 0.48$\pm$0.01 & 0.009$\pm$0.003 & 0.009$\pm$0.001 \\
  Modified poles  & 3.7 & 3.2 & 0.85$\pm$0.03 & \textbf{0.53$\pm$0.02} & 0.009$\pm$0.003 & \textbf{0.012$\pm$0.002} \\
  Modified, $\pm45\degr$  & 3.7 & 3.2 & 0.85$\pm$0.04 & \textbf{0.56$\pm$0.01} & 0.009$\pm$0.002 & \textbf{0.011$\pm$0.001} \\
  Modified, field $\times$2 & 3.7 & 3.2 & \textbf{0.90$\pm$0.01} & 0.49$\pm$0.01 & 0.012$\pm$0.004 & \textbf{0.011$\pm$0.001} \\
\hline
\end{tabular}
\begin{tablenotes}
    \item \textsuperscript{a}In units of km\textsuperscript{-1}~s.
    \item \textit{Note.} For each WSA ($R_{ss}$, $R_i$) pair, those metrics that improve upon the unmodified ADAPT maps by at least one standard deviation are in bold. \label{tab:metrics}
\end{tablenotes}
\end{threeparttable}
\end{table*}

To quantify the agreement between the in-situ observations and WSA SW predictions, \citet{Meadors2020} introduced the WSA prediction \textit{H} metric. This incorporates both the magnetic field polarity and SW speed predictions and compares the discrepancy between their predicted and observed values at a single observatory. It is defined:
\begin{equation}
    \label{eqn:WSA:H}
    H=\frac{\langle\mu\rangle}{\langle\Delta v_{rms}\rangle}
\end{equation}
where $\langle\Delta v_{rms}\rangle$ is the average root-mean-square velocity difference calculated over a fixed window. Since the magnetic field polarity is discrete ($+1$, $-1$, or $0$ if it is indeterminate), rather than the average, $\langle\mu\rangle$ is actually the fraction of correct predictions (where $0$ is considered half-right) over the window. We calculate \textit{H} using a running 7-day window and take the average (and standard deviation) over the entire prediction associated with CR1901. To reduce this comparison to a single number per ensemble, rather than calculating \textit{H} individually for the predictions from each realization, we instead calculate it for the ensemble average speed and polarity at each time step.

These metrics for the predictions at WIND and Ulysses for each ADAPT ensemble are presented in Table \ref{tab:metrics}. This table also includes the M and TSS of each ensemble CH prediction averaged over CR1901. It should be noted that while these metrics represent ensemble averages through time, it is not typically meaningful to use the entire ADAPT ensemble. Instead, commonly only the single ADAPT realization that provides the most reliable coronal field and SW solution for the time and location of interest is used to drive analyses of downstream effects. Nevertheless, those ensembles with the best average performance should contain the most consistently reliable individual realizations (according to a particular metric) unless the ensemble spread is quite large. With that in mind, the ensemble averaged metrics over CR1901 provide some meaningful insights.

Consistent with \citet{Meadors2020}, the solutions using ($R_{ss}$,~$R_i$) = (3.7~$R_\odot$, 3.2~$R_\odot$) provide uniformly better SW predictions at WIND using each set of ADAPT inputs than do the solutions using ($R_{ss}$,~$R_i$) = (2.51~$R_\odot$, 2.49~$R_\odot$). In addition, the larger solution radii improve M. However, using these larger radii yield CHs with reduced TSS and worse predictions of Ulysses observations. This highlights the danger of optimising a complex model such as WSA using only a single, spatially localized metric and suggests that using the larger WSA radii during this period does not yield ``universally'' improved models.

We also see that the polar flux modification typically improves the metrics and never significantly worsens any. In general, modifying the polar fields tends to increase the CH TSS, so the models more correctly capture reality, while only those ADAPT maps with double the polar field strength lead to meaningfully increased M indicating that the prediction in a particular pixel is more likely to be correct. Additionally, each modification consistently improves the predictions at Ulysses, but only the doubling of the polar field strength improves the predictions at WIND. In general then, doubling the polar field strengths has the most consistently positive impact on all the metrics, particularly for the smaller radii WSA solutions, although the improvements are not always largest.

\subsection{WSA solar wind speed calibration}
\label{sec:WSA:solar_wind}
The improvements in the SW speed predictions at Ulysses (sections \ref{sec:WSA:Ulysses} and \ref{sec:WSA:metrics}) are the result of the removal of the spatial structure in the photospheric magnetic field at the poles. Consistent with this, various magnetohydrodynamic (MHD) coronal models \citep[e.g.;][]{Linker2013, Linker2017, Caplan2021} apply smoothing to their driving magnetic field maps, including at the poles, to facilitate computational stability. In addition, a number of authors have used deliberately low-resolution methods to fill in gaps in polar observations \citep{Sun2011, Linker2013, Linker2017}, although recent efforts incorporate methods to introduce realistic flux concentrations \citep{Mikic2018}. However, any smoothing of the magnetic field maps is inconsistent with observations of the polar magnetic field from the Hinode Solar Optical Telescope Spectro-Polarimeter. \citet{Tsuneta2008b} and \citet{Petrie2017} find highly structured polar fields throughout the solar cycle, with kilogauss flux concentrations scattered amongst mixed polarity weak fields, analogous to the low-latitude quiet-Sun. The original ADAPT maps have polar flux concentrations with densities up to only $\sim100$~G but, due to the limited $1\degr$ resolution, these could represent kilogauss flux concentrations with a $\sim10\%$ filling factor. Yet even though the polar fields are already under-resolved in ADAPT, further smoothing resulting from the polar flux modification yields improved SW speed predictions at the cost of ensemble spread.

However, the WSA wind speed prediction (equation \ref{eqn:WSA:speed}) is calibrated using observations in the ecliptic \citep{Wang1990,Arge2000, Arge2003,McGregor2008}. As such, it is perhaps unsurprising that WSA is not able to reproduce the highest wind speeds originating from deep within polar CHs that are only rarely observed in the ecliptic. Even with the improved predictions associated with the modified polar flux demonstrated in Figure \ref{fig:Ulysses:timeseries} and table \ref{tab:metrics}, WSA under-predicts the observed wind speed at Ulysses by $\sim50$~km~s\textsuperscript{-1}. As mentioned in section \ref{sec:WSA:Ulysses}, this is worsened to a $\sim85$~km~s\textsuperscript{-1} discrepancy when using the ($R_{ss}$,~$R_i$) = (3.7~$R_\odot$,~3.2 $R_\odot$) solution radii necessary to achieve the reliable solutions at WIND (section \ref{sec:WSA:WIND}). In reality, a single dynamic corona is responsible for the wind observed at both satellites.

These consistently underpredicted wind speeds at Ulysses may be the result of the WSA SW prediction itself rather than the photospheric magnetic field maps that drive it. Deep within the cores of CHs (when $\theta_b$ is large), equation \ref{eqn:WSA:speed} reduces to
\begin{equation}
\label{eqn:WSA:speed_simplified}
v_{sw}\left(f_{ss}\right) = v_0 + \left(\!\frac{v_m}{\left(1+f_{ss}\right)^{c_1}}\!\right)
\end{equation}
with $v_0=285$~km~s\textsuperscript{-1}, $v_m=625$~km~s\textsuperscript{-1}, and $c_1=\frac{2}{9}$. While theoretically this allows wind speeds up to $910$~km~s\textsuperscript{-1} (as $f_{ss}\to0$), purely radial field expansion yields $v_{sw}\left(f_{ss}=1\right)\approx820$~km~s\textsuperscript{-1} while four-times radial expansion leads to $v_{sw}\left(f_{ss}=4\right)\approx725$~km~s\textsuperscript{-1}. Given that the open polar field (and the relatively small low latitude open field) shown in Figures \ref{fig:WIND:derch} and \ref{fig:Ulysses:derch} must expand to fill the entire sphere at the source surface, it is not surprising that, even deep within the polar CHs, the field expansion is significantly super-radial. In addition, the smaller CHs resulting both from the ADAPT ensembles with double the polar field strength and WSA solutions with ($R_{ss}$,~$R_i$) = (3.7~$R_\odot$, 3.2~$R_\odot$) require more expansion to fill the sphere, leading to further reduced wind speed predictions.

Previous work by \citet{McGregor2011} found a different wind speed parameterization of equation \ref{eqn:WSA:speed} (with $v_0=200$~km~s\textsuperscript{-1}, $v_m=750$~km~s\textsuperscript{-1}, $c_3=3.8$, and $c_4=3.6$) capable of achieving higher SW speeds. This parameterization was calibrated to SW observations from Helios \citep{Schwenn1975} perihelion passes (0.3--0.4~AU) in which the SW experienced minimal processing. This parameterization was then used to drive ENLIL \citep[a 3-D Magnetohydrodynamic heliospheric model;][]{Odstrcil2003} predictions that were found to agree well both at Earth and with Ulysses measurements up to $\sim800$~km~s\textsuperscript{-1} measured at $\pm60\degr$ latitude during a fast latitude scan. Unfortunately, this parameterization cannot be used with the default ballistic WSA SW propagation, yielding SW speed predictions significantly slower than observations for all models during most of CR1901. This and other alternate formulations of equation \ref{eqn:WSA:speed} suggest that it should be possible to tune the parameters for better agreement with SW originating from polar CHs, although \citet{Wang1997b} found that calibrating this relationship using only high-latitude Ulysses measurements creates the reverse problem, resulting in too much fast wind near the ecliptic.

It should be noted that work by \citet{Riley2015} suggests that the primary parameter describing the SW acceleration is the CH boundary distance ($\theta_{b}$) and that the WSA SW velocity calculation may improve if $f_{ss}$ is ignored entirely. This is somewhat in conflict with our findings in section \ref{sec:WSA:Ulysses}. During this period, the SW speed at Ulysses varies by $\pm17$~km~s\textsuperscript{-1} while the modeled spacecraft connectivity never approaches the CH boundary. The predicted wind speed from the unmodified ensemble captures this level of variability (panels (a) and (b), Figure \ref{fig:Ulysses:timeseries}), but the predictions from the ensemble with modified flux (panels (c) and (d)) do not. The flux randomization process homogenizes the field, essentially eliminating intermediate scale spatial variability in the field that causes variations in $f_{ss}$. This implies that a wind speed parameterization based purely on the CH boundary distance would be unable to account for differences in wind speed from within CH interiors. In addition, \citet{Wang2010} found that the average high-latitude SW speed observed by Ulysses depends on the average expansion factor, measuring $\langle v_{sw}\rangle=763$~km~s\textsuperscript{-1} and $\langle f_{ss}\rangle=4.2$ during the solar minimum following solar cycle 22 and $\langle v_{sw}\rangle=740$~km~s\textsuperscript{-1} and $\langle f_{ss}\rangle=5.1$ during the solar minimum following solar cycle 23. It is not clear that a SW speed based solely on the CH boundary distance can explain this result.

\section{Conclusion}
\label{sec:conclusion}
In this work we investigated the importance of polar magnetic fields on coronal magnetic field models and the prediction of the SW throughout the heliosphere. For this study we focused on Carrington rotation 1901, a period that was recently investigated by \citet{Meadors2020} to optimize WSA solutions and during which the Ulysses spacecraft was embedded in the SW at high heliospheric latitudes. We modified the polar photospheric magnetic field of ADAPT global magnetic maps, using a novel flux randomization procedure to closely align with observed CH boundaries, and used WSA to predict CHs and the SW at the WIND and Ulysses spacecraft.

Consistent with \citet{Meadors2020}, we find that the interface and source surface radii used by WSA significantly impact combined PFSS and SCS coronal models. Moderate adjustments to the selected radii values result in large changes in the modeled CHs and SW predictions independent of the driving photospheric magnetic map. In particular, during the CR1901 period, ($R_{ss}$,~$R_i$) = (3.7~$R_\odot$, 3.2~$R_\odot$) yield significantly smaller modeled CHs than the standard ($R_{ss}$,~$R_i$) = (2.51~$R_\odot$, 2.49~$R_\odot$), resulting in more reliable predictions of the CHs (M increases) but correctly identifying less of their area (TSS decreases). The SW predictions at WIND are improved significantly because of changes in the spacecraft connectivity from small, low-latitude CHs to the large polar CHs. However, using the larger solution radii worsens the predictions at Ulysses, decreasing the predicted wind speeds which were already not able to regularly capture the observed $780\pm17$~km~s\textsuperscript{-1} speeds over CR1901. This is a result of the generally smaller polar CHs resulting from the larger WSA solution radii since they necessarily force the field to expand further to fill the heliosphere and the WSA-predicted SW speed is inversely related to the expansion factor.

By modifying the polar flux distribution based on observed CH boundaries, the modeled CHs more closely agree with the observations. None of the polar flux redistribution modifications were found to worsen the predictions according to any metric, and when combined with doubling the polar field strength, the polar modification yields improvements across all four metrics using WSA with ($R_{ss}$,~$R_i$) = (2.51~$R_\odot$, 2.49~$R_\odot$). The internal homogeneity imposed by the flux randomization procedure results in generally sharper CH boundaries and often smaller SW speed variation between ADAPT ensemble realizations. This later effect is particularly noticeable in the Ulysses SW speed predictions which become dramatically more consistent between ensemble realizations and also have greater average speeds that are more in line with the observations. This consistent offset between the Ulysses SW speed observations and WSA predictions (even after the flux modification) suggests that the WSA SW speeds may need to be recalibrated for high-latitude wind predictions from polar CHs.

We also find that the large-scale structure of the polar CHs is significantly impacted by the low-latitude flux concentrations (i.e. active regions) in the ADAPT maps. When the low-latitude flux concentrations are dispersed, the modeled CHs become much more responsive to the polar flux redistribution procedure. This highlights limitations with current global solar magnetic maps due to the lack of farside magnetogram observations and difficulties associated with flux assimilation in near real time. It also suggests that modifying the polar field to conform to observed CH boundaries could provide a valuable diagnostic of low latitudes on the farside of magnetic maps.

The consequences of limited magnetograph observational views are particularly important in light of the successful launch of the Polarimetric and Helioseismic Imager \citep[PHI;][]{Solanki2020} aboard Solar Orbiter \citep[SolO;][]{Muller2020}. PHI will soon provide photospheric magnetic field observations of the farside and, for the first time, the solar poles from above the ecliptic. However, we now anticipate that both of these new SolO/PHI perspectives will be needed simultaneously to provide substantially improved global maps of the photospheric magnetic field. Combining nearside and farside full-disk magnetograms with vector measurements at the poles \citep[e.g. figures 10 and 11 of][]{Petrie2017} will provide the most comprehensive instantaneous measurement of the global photospheric magnetic field to date. Comparing derived open magnetic fields from these maps with dynamic CH observations \citep[e.g.,][]{Caplan2016} will allow for continuous validation of both the ADAPT and WSA models and improved characterization of the corona and SW.

\begin{acknowledgments}
This work utilizes data produced collaboratively between Air Force Research Laboratory (AFRL) and the National Solar Observatory (NSO). The ADAPT model development is supported by AFRL, along with AFOSR (Air Force Office of Scientific Research) tasks 18RVCOR126 and 22RVCOR012. The views expressed are those of the authors and do not reflect the official guidance or position of the United States Government, the Department of Defense or of the United States Air Force. The input data utilized by ADAPT is obtained by NSO/NISP (NSO Integrated Synoptic Program). NSO is operated by the Association of Universities for Research in Astronomy (AURA), Inc., under a cooperative agreement with the National Science Foundation (NSF). NSO/Kitt Peak data used here are produced cooperatively by NSF/NOAO, NASA GSFC, and NOAA/SEL. The coronal hole data used here were compiled by K. Harvey and F. Recely using NSO KPVT obseravtions under a grant from the NSF. SJS would like to thank Samantha Wallace for providing an image version of the \citet{Harvey2002} coronal hole map. The WIND data distributed with WSA were originally processed by T. R. Detman, and the full WIND and Ulysses data archives are available online at \url{https://omniweb.gsfc.nasa.gov/}. C. Nick Arge is supported by the NASA competed Internal Scientist Funding Model (ISFM). The data used for the analyses in this paper and additional data products are available online at \url{https://doi.org/10.5281/zenodo.6309825}.
\end{acknowledgments}

%

\facilities{NSO(KPVT), WIND(SWE and MFI), Ulysses(SWOOPS and MFI)}


\software{WSA \citep[v5.3.2][]{Arge2000, Arge2003, Arge2004a},
          SunPy \citep[v2.0.6][available online at \url{https://doi.org/10.5281/zenodo.4421322}]{SunPy2020},
          NumPy \citep[v1.19.4][]{Harris2020},
          Pandas \citep[v1.1.5][available online at \url{https://doi.org/10.5281/zenodo.4309786}]{McKinney2010},
          Matplotlib \citep[v3.3.3][available online at \url{https://doi.org/10.5281/zenodo.4268928}]{Hunter2007},
          SciPy \citep[v1.5.3][]{Virtanen2020a},
          Astropy \citep[v4.2][]{Robitaille2013, Price-Whelan2018}
          }



\appendix

\section{Polar Flux Redistribution Algorithm}
\label{sec:appendix:randomization_algorithm}
Randomizing the flux in the polar regions of global solar magnetic maps is accomplished in two phases:
\begin{enumerate}
    \item Fluxes above the cutoff latitude (Section \ref{sec:ADAPT:cutoff}) from the original map are sorted into arrays to fill the coronal hole (CH) and non-CH areas.
    \begin{itemize}
        \item Calculate the number of pixels of each polarity needed to fill the output map based on the observed CHs and prescribed percent unipolarity (by number of pixels).
        \item For each polarity, fluxes (converted from flux densities) are resampled from the input magnetogram.
        \begin{itemize}
            \item If the input has more pixels than needed: the output distribution is randomly drawn without replacement from the input.
            \item If the input has fewer pixels than needed: the input distribution is copied a whole number of times, with the remainder randomly drawn without replacement from the input.
        \end{itemize}
        \item Each polarity distribution is rescaled to conserve the positive, negative, signed, and unsigned flux.
        \item The distributions of each polarity are randomly combined into two arrays to fill the CH and non-CH regions according to the prescribed unipolarity.
    \end{itemize}
    \item The arrays are used to populate the CH and non-CH areas of the output map separately.
    \begin{itemize}
        \item In absolute descending order, each flux of the appropriate array is randomly assigned to an empty pixel in the associated region, with probabilities weighted by the pixel area (such that larger fluxes are preferentially assigned to larger pixels). Fluxes are only placed into pixels to yield flux densities less than that of the largest flux placed into the largest pixel.
        \item The redistributed fluxes are converted back into flux densities and inserted into the original map.
    \end{itemize}
\end{enumerate}
The restriction about not placing large fluxes into small pixels ensures that this process does not create physically unrealistic large flux densities by assigning a large flux from originally larger area pixels (i.e., lower latitude) into smaller area pixels at higher latitudes.

\section{Polar Flux Scaling Algorithm}
\label{sec:appendix:scaling_algorithm}
\begin{figure*}[!t]
    \includegraphics[width=\textwidth]{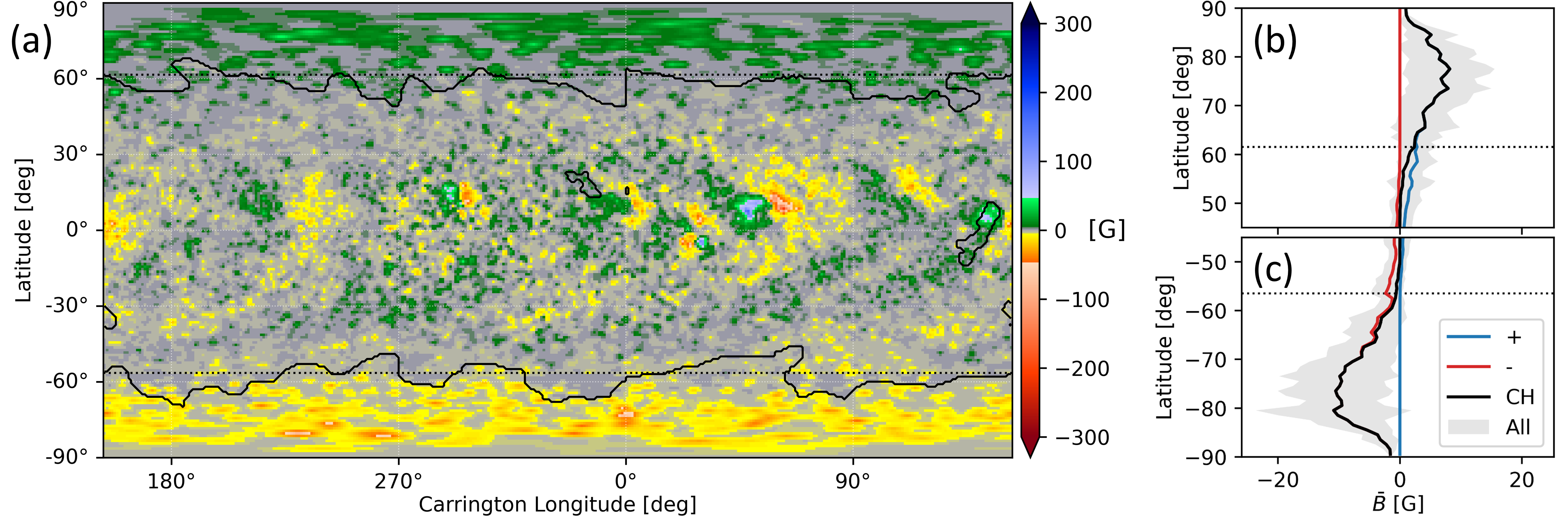}
    \caption{The first realization of the ADAPT ensemble on 1995 October 1\textsuperscript{st} for the polar scaling field modification. Panel (a) shows the radial photospheric magnetic field and the solid black contours mark the observed coronal hole (CH) boundaries from \protect\citet{Harvey2002}. Panels (b) and (c) indicate the average magnetic field strength as a function of latitude in the north and south poles, respectively. They show the average of the positive (blue), negative (red), and CH (black) flux along with the standard deviation of all fluxes (gray shaded region). In all panels, the horizontal dotted black lines indicate the latitude boundaries set by the ADAPT data assimilation range of $-56\degr$ to $+61\degr$ between which the flux is unmodified.
    \label{fig:ADAPT_scaled}}
\end{figure*}

To force the ADAPT map poles to conform to the observed coronal hole (CH) boundaries while preserving the spatial structure of the magnetic field created by the ADAPT map flux transport, we implemented a modification scheme based around changing the magnetic field zero point. By subtracting a small constant flux density from pixels in the polar regions, we changed the percent unipolarity, and then the positive, negative, net, and total flux are conserved by multiplicatively scaling the positive and negative flux density independently. An example ADAPT map resulting from this polar flux scaling process is shown in Figure~\ref{fig:ADAPT_scaled}. 

Comparing the example ADAPT realization map in Figure~\ref{fig:ADAPT_scaled} with the unmodified ADAPT map in Figure \ref{fig:ADAPT_original} panel (a), it is clear that this flux scaling modification yields very small changes, particularly compared to those resulting from the polar flux randomization. None of the ensemble and CR1901 averaged metrics showed improvements statistically greater than those from the equivalent flux-randomized ensembles, and the TSS and \textit{H}(Ulysses) are found to be statistically worse. Consequently, despite the more physically realistic spatial structuring of the polar fields resulting from this procedure, we focused our efforts on the polar randomization technique (outlined in Section \ref{sec:ADAPT:modification} and Appendix \ref{sec:appendix:randomization_algorithm}).

The procedure for the polar flux scaling algorithm is as follows:
\begin{itemize}
    \item Above the cutoff latitude, identify the two regions inside and outside the observed CH.
    \item For each region that does not meet the minimum unipolarity threshold:
    \begin{itemize}
        \item Determine the constant flux density offset necessary to achieve the minimum unipolarity threshold by:
        \begin{itemize}
            \item Ordering the pixels in the region by their flux density, starting with the maximum flux density of the dominant polarity.
            \item Cumulatively sum the areas of the pixels.
            \item Average the flux densities of the first two pixels with cumulative areas exceeding the unipolarity threshold.
        \end{itemize}
        \item Add this offset value to all pixels in the region.
    \end{itemize}
    \item If the offset is added to either region, then:
    \begin{itemize}
        \item Re-scale all positive flux densities above the cutoff latitude by the ratio of the original total positive flux to the modified total positive flux, ensuring the total positive flux is unchanged.
        \item Repeat for the negative fluxes.
    \end{itemize} 
\end{itemize}
The final step ensures that the positive and negative (and therefore net and total) flux is conserved in the re-scaling process. It is also possible for the original ADAPT map to satisfy the applied unipolarity thresholds in each pole, in which case the map is not modified. This condition was not met for any of the ADAPT maps used in this study.

\section{WSA Model Run Parameters}
\label{sec:appendix:parameters}
\begin{table*}[!th]
\caption{WSA parameters used for all model run results in this paper}
\centering
\begin{tabular}{l c l}
\hline
  Tag    &   Value   &   Description \\
\hline
  MAPTYPE               &   DU      &   Model operation mode, daily updated magnetograms \\
  GRID                  &   2.0     &   Latitude and longitude spatial resolution of the coronal model in degrees\\
  SUBSAT\_OFFSET        &   1.0     &   $\pm$ latitude offset in degrees to determine uncertainty at satellite \\
  OUTER\_RAD            &   5.0     &   Outer radius of model volume in $R_\odot$ \\
  NM\_SPHAR             &   90      &   Number of spherical harmonics used to calculate the PFSS model \\
  DS\_PFSS\_IN          &   -0.01   &   Inward field line tracing step size through the inner PFSS shell \\
  DS\_SCS\_IN           &   -0.05   &   Inward field line tracing step size through the outer SCS shell \\
  DS\_PFSS\_OUT         &   0.01    &   Outward field line tracing step size through the inner PFSS shell \\
  DS\_SCS\_OUT          &   0.05    &   Outward field line tracing step size through the outer SCS shell \\
  RUNCURRENTSHEET       &   1       &   Binary toggle set to run with SCS model \\
  DEL\_DAY              &   0.5     &   Window in days to determine the solar wind prediction from each input map \\
  MAXFILL               &   0       &   Number of days to predict to fill in for missing input files \\
  PARAMETERIZE\_VEL\_RELS   &   1       &   Binary toggle set to use the velocity equation \\
  MONOPOLE\_CORR         &   2       &   Type of monopole correction: scaled \\
\hline
\label{tab:parameters}
\end{tabular}
\end{table*}

Table \ref{tab:parameters} contains the WSA model parameters used for all the runs in this paper.

\section{WSA Model Inner Boundary}
\label{sec:appendix:inner_boundary}
\begin{figure*}[!t]
    \includegraphics[width=\textwidth]{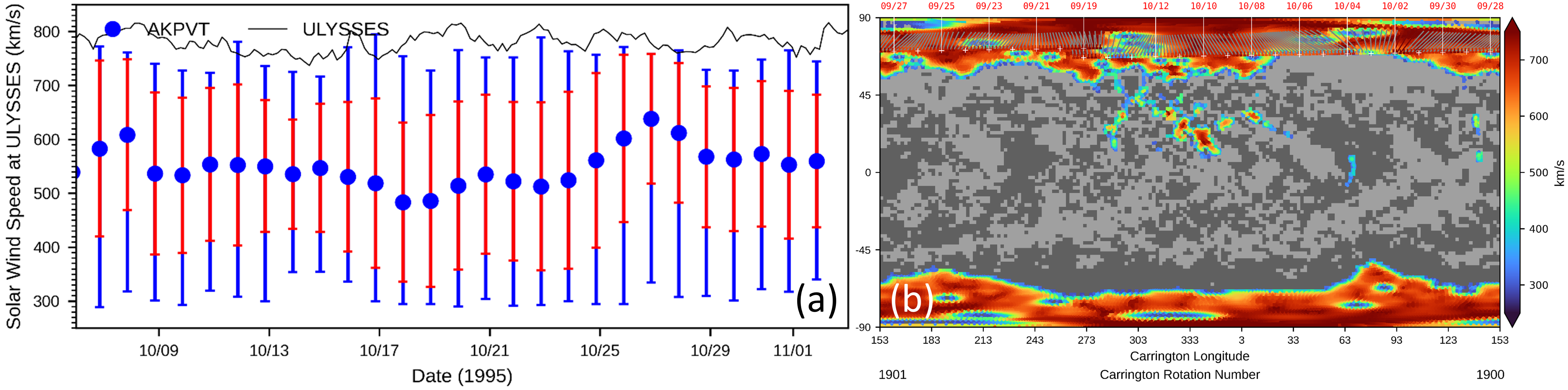}
    \caption{For WSA 5.2 with open field traced from the photosphere (R=1.0~$R_\odot$) using the unmodified ADAPT maps. Panel (a) shows the 6-day solar wind (SW) speed predictions at the Ulysses spacecraft over CR1901 compared with observations (with the same formatting as Figures \ref{fig:WIND:timeseries} and \ref{fig:Ulysses:timeseries}) and panel (b) shows the WSA predicted coronal holes, SW speed, and Ulysses spacecraft connectivity from the first realization of the ADAPT ensemble on 1995 October 1\textsuperscript{st} (with the same formatting as Figures \ref{fig:WIND:derch} and \ref{fig:Ulysses:derch}).
   \label{fig:Ulysses:old_tracing}}
\end{figure*}

The WSA model estimates the open photospheric magnetic field in two ways: by tracing the magnetic field lines 1) from the grid of open field on the outer boundary down to the photosphere and 2) from the photosphere up to the outer boundary. In WSA version 4.3 (beginning on 2015 August 25\textsuperscript{th}), the field line tracing from the photosphere outward changed from starting at a user-defined radius, typically 1.01~$R_\odot$, to exactly the input boundary at 1.0~$R_\odot$. This is physically reasonable since that is the radius of the observed photospheric magnetic field, but doing so has dramatic negative impacts on the resulting WSA solutions. For example, compare Figure \ref{fig:Ulysses:old_tracing} generated by tracing from 1.0~$R_\odot$ to panel (a) of Figures \ref{fig:Ulysses:timeseries} and \ref{fig:Ulysses:derch}, both generated by starting the tracing at 1.01~$R_\odot$. Starting the in-to-out line tracing from the photosphere results in a number of closed-field regions embedded within the polar open fields, indicated by the large patches of slow solar wind (SW) from the interiors of the polar coronal holes (CHs) in panel (b). These closed field regions have a dramatic negative impact on SW speed predictions due to its dependence on the distance to the CH boundary in equation \ref{eqn:WSA:speed}. As a result, the ensemble average decreases by $\sim150$~km~s\textsuperscript{-1} and the variability within the ensemble increases substantially as seen in panel (a), essentially encompassing the observable range of SW speeds at all times.

The closed field regions at the poles when using the lower tracing start height are a result of numerical artifacts in the spherical harmonic expansion used to extrapolate the PFSS itself. These are similar to numeric ringing typical of Fourier Transform-like representations of sharp boundaries. At the photosphere, in regions where the field is weak, small numerical errors in the harmonic expansion can cause the field polarity to flip in the model compared to the original input observation (i.e., the values in the photospheric field map used to drive the model). Even though these effects might be quite small on an absolute scale, the change in sign dramatically influences the resulting magnetic field. This is particularly problematic near the photosphere where the field has the most fine-scale spatial structure. By 1.01~$R_\odot$ (the next radial layer in the coronal model), the field is significantly simpler and these sign changes no longer occur. Consequently, tracing the field from this first layer in the model results in significantly more realistic polar CHs. WSA version 5.3.2 (released in 2021 November) restores the functionality that existed prior to version 4.3, now always tracing the field outward starting at 1.01~$R_\odot$.

Besides avoiding numerical artifacts, there is also physical justification for starting the field tracing above the photosphere. The PFSS extrapolation of the coronal magnetic field near the Sun assumes that the domain has no electric currents and the plasma $\beta$ is small. However, this condition is not met in the photosphere which serves as the boundary of the extrapolation. The low-$\beta$ assumption is a general problem with all force-free-field extrapolations and makes the magnetic connectivity near the photosphere particularly suspect. In reality, the polar field has mixed polarities and there are small pockets of closed fields at the poles. However, these opposite polarities quickly close in the chromosphere, forming the magnetic canopy at a height of about 1 Mm \citep[e.g.,][and references therein]{Wiegelmann2014}. In CHs with a strong polarity imbalance, above the canopy the coronal field should be essentially unipolar. Starting the field tracing at 1.01~$R_\odot$ ($\sim7$~Mm) is already well into the corona and avoids these difficulties associated with the chromosphere. In addition, the majority of SW acceleration occurs above r=1.5~$R_\odot$ \citep[and references therein]{Cranmer2009}, so the small scale properties of the photospheric magnetic field may not be essential model input.


\bibliography{references}{}
\bibliographystyle{aasjournal}



\end{document}